\documentclass[aps,pre,reprint,superscriptaddress,longbibliography,floatfix]{revtex4-2}
\usepackage[T1]{fontenc}
\usepackage[utf8]{inputenc}
\usepackage{lmodern}
\usepackage{amsmath,amssymb,bm}
\usepackage{graphicx,booktabs}
\usepackage{microtype}
\usepackage[hidelinks]{hyperref}

\newcommand{\E}{\mathrm E}
\newcommand{\Var}{\operatorname{Var}}
\newcommand{\Lgen}{\mathcal L}
\newcommand{\Dresp}{\mathcal D}

\begin{document}

\title{Attributing extreme-event probability to a source variable}

\author{Daniel F. T. Hagan}
\email{daniel.hagan@ugent.be}
\affiliation{Hydro-Climate Extremes Lab (H-CEL), Ghent University, Ghent, Belgium}


\begin{abstract}
Information-flow theory quantifies directional coupling through the rate of change of a target's Shannon entropy, a bulk functional that does not identify the probability of any particular threshold event. Here, we target extreme events by asking how a source variable contributes to the probability that the target exceeds a threshold. For source-additive drift the source's share of the marginal probability current is exact and separates into a mean-forcing part and a conditional-excess part that vanishes under independence. An identity links the two descriptions: the Liang information flow is the density-weighted mean of the derivative of the specific source current, whereas the exceedance current is its level at the threshold. Entropy-based coupling therefore collapses in saturated regimes, where the source supplies its largest forcing but that forcing no longer responds to the source, and R\'enyi reweighting, higher-order expansion and Fisher normalisation are suppressed for the same reason. The exact flux decomposition does not attribute either, because its terms are gross transports that nearly cancel. An adjoint response built from the backward generator does, once the reference level of the source is stated. For a perturbation that removes the source's conditional excess, its relative reading falls short of the true effect by about a quarter, nearly independently of event probability over two decades, whereas two of the three misspecifications we test bias it upward, by up to a factor of two. When the perturbation instead shifts the target's mean, the shortfall has a closed form and shrinks toward the tail. Applied to the 2003 European and 2010 Russian heatwaves in reanalysis, the attributed risk ratio rises with threshold from near one to about two at the 2003 site and to an order of magnitude at the 2010 site, where the fitted model is extrapolated. Both curves follow from the mean warming and the predictive spread alone, so a contribution quoted as a ratio without its threshold is underspecified.
\end{abstract}

\maketitle

\section{Introduction}

Whether a source variable contributes to an extreme event is a question about probability, and not about entropy. Yet the directional-coupling measures in widest use in the physical sciences are entropy rates. The Liang--Kleeman construction defines information transfer by comparing the evolution of a target's marginal Shannon entropy with its evolution when a source is frozen \cite{LiangKleeman2005,Liang2008,Liang2016}, and linear \cite{Liang2014} and nonlinear \cite{Pires2024} estimators have made it usable on observed series, including in climate applications where the events of interest are extremes \cite{Vannitsem2025,Hagan2019,Zhou2024,Zhang2026}.

\begin{figure*}[t]
\includegraphics[width=\textwidth]{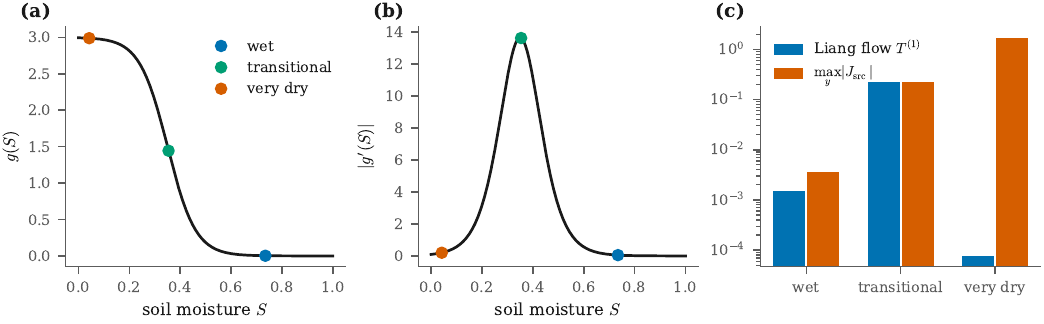}
\caption{Level and slope of the source response are different quantities, and only the level tracks transport across a threshold. (a) The saturating response $g(S)$ of the testbed, with the mean soil-moisture state of the three equilibria marked. (b) Its derivative $|g'(S)|$, the sensitivity of the forcing to variations of the source, which collapses in both saturated limits and takes with it the slope of the conditional source mean that the entropy transfer weights. (c) Liang information flow and the largest source current $\max_y|J_{\rm src}|$ in each regime, logarithmic scale, with the source forcing measured from its wet-limit value $g=0$. In the very dry regime the information flow is three orders of magnitude below its transitional value while the source current is nearly eight times larger, because the response is at its maximum level and its minimum slope.}
\label{fig:levelslope}
\end{figure*}

Since Shannon entropy is a bulk functional, it is dominated by the body of the distribution, and an entropy rate does not identify the probability of any particular threshold event. A pure translation of a distribution, for instance, changes every exceedance probability but leaves every differential R\'enyi entropy unchanged. The mismatch is not subtle. In the saturating soil-moisture model used as a testbed below, the very dry regime makes a temperature above $33\,^{\circ}$C almost nine hundred times as likely as the transitional regime does, because its soil forcing is at its maximum. Yet its information flow from soil to temperature is only $7.7\times 10^{-5}$ nats per day, three thousand times less than in the transitional regime, since at saturation that forcing no longer varies with soil moisture [Fig.~\ref{fig:levelslope}]. This is not an estimator failure, but rather an indication that the diagnostic is correctly reporting a quantity that was never about the tail.

This paper asks the tail question directly. Section~\ref{sec:flux} constructs the exact decomposition of the rate of change of exceedance probability into contributions labelled by the terms of the drift. It establishes the properties that make the source term interpretable as a coupling diagnostic, relates it to the Liang information flow through an identity, and shows that no available repair of that measure escapes the distinction the identity draws. Section~\ref{sec:nonattr} shows that the exact decomposition nevertheless fails to attribute, and why. Section~\ref{sec:adjoint} gives the object that does attribute, an adjoint response computed from the backward generator without counterfactual integration, identifies the functional in which its first-order error depends least on the rarity of the event, and gives that error in closed form when the perturbation shifts the target's mean. Sections~\ref{sec:estimation} and~\ref{sec:numerics} give the estimator and map its operating envelope against a ground-truth counterfactual, including the saturated regime and the role of the reference against which a contribution is measured, and Sec.~\ref{sec:application} applies it to two observed heatwaves.

Two features of the construction are worth stating at the outset. First, the decomposition needs no freezing operation, because it is an algebraic split of a current whose terms already carry the labels of the physical drift. The attribution rests on one, and Sec.~\ref{sec:estimation} identifies it and explains why it is placed differently from the freeze of Liang \cite{Liang2016}. Second, the attribution is computed from the fitted generator alone. Counterfactual attribution of extremes conventionally requires running the counterfactual, with large ensembles or prescribed-forcing experiments from which risk ratios are formed \cite{Stott2016}, and counterfactual causal theory has made precise what such a ratio means \cite{Hannart2016}. That route is unavailable for internal feedbacks at scale, and for observations altogether. Linear response theory offers an alternative that needs no counterfactual run, and it has recently been given a rigorous footing for the detection and attribution of forced change in the mean state \cite{Lucarini2024}. Here the same kind of response is computed for the probability of an exceedance.

\section{Exceedance flux and its decomposition}
\label{sec:flux}

\subsection{The marginal probability current}

Consider the It\^o system
\begin{equation}
 d\bm U = \bm F(\bm U,t)\,dt + \bm B(\bm U,t)\,d\bm W,
 \qquad \bm D = \bm B\bm B^{\top},
 \label{eq:sde}
\end{equation}
with target $Y$, source $X$ and remaining coordinates $\bm Z$. Integrating the Fokker--Planck equation over $(X,\bm Z)$ gives the exact evolution of the target marginal \cite{Risken1989,Gardiner2009},
\begin{equation}
 \partial_t\rho = -\partial_y\!\left[\bar F(y,t)\rho\right]
 + \tfrac12\,\partial_y^2\!\left[\bar D(y,t)\rho\right],
 \label{eq:marginalfp}
\end{equation}
where $\rho(y,t)$ is the marginal density of $Y$ and
\begin{equation}
 \bar F(y,t)=\E[F_Y\mid Y=y],\qquad
 \bar D(y,t)=\E[D_{YY}\mid Y=y]
 \label{eq:condcoef}
\end{equation}
are conditional expectations of the drift and diffusion coefficients. No closure is invoked here. Equation~\eqref{eq:marginalfp} is exact, and the price of marginalisation is that its coefficients are conditional rather than pointwise. Defining the marginal probability current
\begin{equation}
 J(y,t) = \bar F(y,t)\rho(y,t)
 - \tfrac12\,\partial_y\!\left[\bar D(y,t)\rho(y,t)\right]
 \label{eq:current}
\end{equation}
puts Eq.~\eqref{eq:marginalfp} in the conservation form
$\partial_t\rho = -\partial_y J$.

\subsection{The exceedance-flux identity}

Fix a threshold $u$ and let $P_u(t)=\int_u^{\infty}\rho(y,t)\,dy$. Provided
$J(y,t)\to0$ as $y\to\infty$, integrating the conservation form over $[u,\infty)$
gives
\begin{equation}
 \boxed{\;\frac{dP_u}{dt} = J(u,t)\;}
 \label{eq:T1}
\end{equation}
The rate of change of exceedance probability is the probability current evaluated at the threshold. Equation~\eqref{eq:T1} is exact for any drift and diffusion satisfying the stated decay, and it becomes a statement about coupling once $\bar F$ is decomposed.

Restrict the target drift to the source-additive class
\begin{equation}
 F_Y = f(Y,\bm Z,t) + b_t X,
 \label{eq:additive}
\end{equation}
which is the class for which the nonlinear Liang transfer \cite{Pires2024}
reduces to a conditional-mean representation (Appendix~\ref{app:bridge}). Writing
$m_t(y)=\E[X\mid Y=y]$ and $\bar f(y,t)=\E[f\mid Y=y]$, the tower property gives
$\bar F = \bar f + b_t m_t$, so the current separates into three terms carrying
the labels of the drift:
\begin{align}
 J &= J_{\rm self} + J_{\rm src} + J_{\rm diff},\label{eq:split}\\
 J_{\rm self}(y,t) &= \bar f(y,t)\rho(y,t),\nonumber\\
 J_{\rm src}(y,t) &= b_t\,m_t(y)\,\rho(y,t),\nonumber\\
 J_{\rm diff}(y,t) &= -\tfrac12\,\partial_y[\bar D\rho].\nonumber
\end{align}
The decomposition requires no freezing construction. It is an algebraic separation of Eq.~\eqref{eq:current} induced by the additive structure of the drift, and each term is separately computable from quantities that a fitted drift model already supplies, namely the coefficient $b_t$, the conditional source mean $m_t$ and the target density $\rho_t$.

\subsection{Mean forcing and conditional excess}

The source term of Eq.~\eqref{eq:split} is not yet a coupling diagnostic. A source that is statistically independent of the target but has nonzero mean still transports probability across $u$, because its mean contributes $b_t\E[X]$ to the drift everywhere. Separating that part,
\begin{equation}
 J_{\rm src} = \underbrace{b_t\,\E_t[X]\,\rho_t(y)}_{\textstyle J_{\rm mean}}
 \;+\;
 \underbrace{b_t\!\left(m_t(y)-\E_t[X]\right)\rho_t(y)}_{\textstyle J_{\rm exc}},
 \label{eq:T2}
\end{equation}
gives a mean-forcing current, present whenever the source has nonzero mean whether or not it is coupled, and a conditional-excess current measuring the additional transport attributable to the source because the target is at $y$. Since $X\perp Y$ implies $m_t\equiv\E_t[X]$, the conditional-excess current vanishes identically under independence, which is the analogue for this functional of the nil-causality property of the entropy transfer \cite{Liang2016}. It is also unchanged if a constant is moved between $f$ and $b_tX$ without changing the total drift, whereas $J_{\rm self}$, $J_{\rm src}$ and $J_{\rm mean}$ are not, so only $J_{\rm exc}$ is independent of how the self and source terms are separated. In the intended application this is the operative split, because it separates the average forcing a source exerts on the target from the additional forcing it supplies specifically in extreme states.

\begin{figure*}[t]
\includegraphics[width=\textwidth]{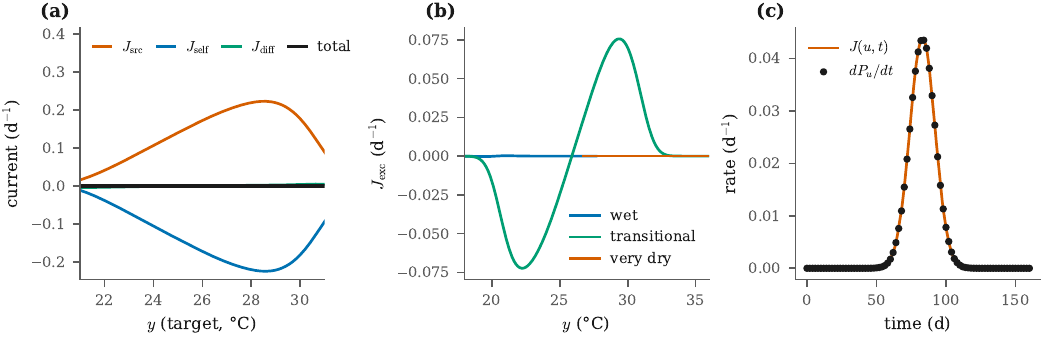}
\caption{The decomposition of Eq.~\eqref{eq:split} and the identity
Eq.~\eqref{eq:T1}. (a) The three current terms in the transitional equilibrium. The total vanishes to numerical precision while the source and self terms are individually of order $0.2$, so a stationary state balances transport rather than suppressing it. (b) The conditional-excess current of Eq.~\eqref{eq:T2} in the three regimes. (c) Verification of $dP_u/dt=J(u,t)$ along a nonstationary traversal at $u=29$. The two curves are indistinguishable at this scale, with a median relative error of $1.9\times10^{-4}$ over five thresholds and seventeen times.}
\label{fig:decomposition}
\end{figure*}

\subsection{Relation to the Liang information flow}
\label{sec:bridge}

The two descriptions are functionals of the same three objects. Define the specific source current $j(y,t)=J_{\rm src}(y,t)/\rho_t(y) = b_t m_t(y)$. Under the drift class \eqref{eq:additive}, with $D_{YY}$ independent of $X$, the local response appearing in the nonlinear Liang--Kleeman information transfer is $\Dresp(y,t) = b_t m'_t(y)$
(Appendix~\ref{app:bridge}), that is
$\Dresp = \partial_y j$, and therefore
\begin{equation}
 \boxed{\;
 T_{X\to Y}(t)
 = \int \partial_y\!\left[\frac{J_{\rm src}(y,t)}{\rho_t(y)}\right]
 \rho_t(y)\,dy \;}
 \label{eq:T3}
\end{equation}
The information flow is the density-weighted mean of the derivative of the specific source current, whereas the exceedance flux is its level at the threshold. Equation~\eqref{eq:T3} makes the distinction structural rather than heuristic, and it explains the saturation behaviour noted in the introduction. Wherever the conditional source mean is locally flat, $\partial_y j$ vanishes and the entropy transfer collapses, while $j$ itself, and hence the transport of probability across a threshold in that region, may be at its largest [Fig.~\ref{fig:levelslope}(c)].

\subsection{Limits of a repaired entropy measure}
\label{sec:repairs}

Equation~\eqref{eq:T3} concerns one measure at one order. Because the tail insensitivity it exposes is a property of the entropy functional rather than of the Liang construction, it is natural to ask whether a modified measure removes the problem. Three repairs are available in the literature, although none changes the conclusion, and the reason is the same in all three.

\subsubsection{Reweighting the entropy}

The R\'enyi entropies are the standard instrument for shifting weight toward the tail. Palu\v{s} and co-workers built a R\'enyi reformulation of the transfer entropy precisely to identify the variable responsible for extreme values of an effect variable \cite{Palus2024}, which makes it the most direct competitor to the construction developed here. Let
$H_\alpha=(1-\alpha)^{-1}\log Z_\alpha$ with $Z_\alpha=\int\rho_t^{\alpha}dy$,
and define the \emph{escort} density
\begin{equation}
 \rho_{\alpha}(y,t)=\rho_t(y)^{\alpha}\big/Z_\alpha(t),
 \label{eq:escort}
\end{equation}
a probability density for every $\alpha>0$ for which $Z_\alpha$ is finite. Differentiating $H_\alpha$ along Eq.~\eqref{eq:marginalfp} and retaining the drift contribution gives
$-\alpha[(1-\alpha)Z_\alpha]^{-1}\int\rho^{\alpha-1}\partial_y[\bar F\rho]\,dy$.
Integrating by parts,
$\int\rho^{\alpha-1}\partial_y[\bar F\rho]\,dy
=-(\alpha-1)\int\rho^{\alpha-1}(\partial_y\rho)\bar F\,dy$, which reduces the
prefactor to $-\alpha/Z_\alpha$. Using
$\rho^{\alpha-1}\partial_y\rho=\alpha^{-1}\partial_y(\rho^{\alpha})$ and
integrating by parts once more leaves
$Z_\alpha^{-1}\!\int\rho^{\alpha}\,\partial_y\bar F\,dy$. Under
Eq.~\eqref{eq:additive} the part of $\partial_y\bar F$ carrying the label of the source is $\partial_y j$, so that
\begin{equation}
 \boxed{\;
 T^{(\alpha)}_{X\to Y}(t)=\int \partial_y j(y,t)\,\rho_{\alpha}(y,t)\,dy\;}
 \label{eq:T4}
\end{equation}
with $\alpha=1$ recovering Eq.~\eqref{eq:T3}. No freezing operation enters, for the same reason it does not enter Eq.~\eqref{eq:split}. We call $T^{(\alpha)}$ the R\'enyi analogue of the Liang flow. It is not the R\'enyi transfer entropy of Ref.~\cite{Palus2024}, which is a lagged difference between conditional R\'enyi entropies of the target's future given its own past with and without the source's past, and the two share neither definition nor units. We use $T^{(\alpha)}$ because it is the population functional that isolates what R\'enyi reweighting does to the response; the lagged statistic is examined numerically in Appendix~\ref{app:renyi}.

The R\'enyi index therefore changes the measure against which the local response is averaged, and nothing else. Since $\rho_\alpha$ is a probability density,
\begin{equation}
 \left|T^{(\alpha)}_{X\to Y}\right|\;\le\;\sup_y\left|\partial_y j(y,t)\right|
 \qquad\text{for every }\alpha,
 \label{eq:bound}
\end{equation}
so the whole family is bounded by the supremum of the slope and never by the level. For Gaussian $\rho=\mathcal N(\mu,s^2)$ one has $\rho_\alpha=\mathcal N(\mu,s^2/\alpha)$. The escort weight broadens as $\alpha\to0$ and contracts onto the mode as $\alpha\to\infty$ while remaining centred at $\mu$ throughout, and there is no $\alpha$ for which it localises at a threshold [Fig.~\ref{fig:repairs}(a)]. This is consistent with the finding that R\'enyi transfer-entropy estimates become unreliable at small $\alpha$ and can invert the ordering of coupling strengths there \cite{Tabachova2026}.

\subsubsection{Expanding in the response time}

A second repair addresses a different way for the flow to vanish. Smirnov \cite{Smirnov2026} observes that for strong bidirectional couplings of a certain antisymmetric kind the stationary density factorises, so the Liang--Kleeman flow is zero in both directions while the coupling coefficients are arbitrarily large. Within the dynamical-causal-effects formalism \cite{Smirnov2022} that flow is the first derivative, with respect to response time, of a finite-time information response. Where the first derivative cancels, the response still rises quadratically, and the second-order differential response, which is proportional to the square of the coupling coefficient, recovers the effect [Fig.~\ref{fig:repairs}(b)].

Here, the two zeros are not the same object. In Smirnov's case the conditional source mean is flat because the coupling produces no correlation between source and target, while the coupling coefficient itself is large, and the second-order response, proportional to its square, recovers it. In the saturated regime the zero of Eq.~\eqref{eq:T3} arises because the physical coupling to variations of the source, $g'(S)$ in the testbed, has itself vanished over the states the source occupies. The second-order response scales as the square of that coupling, so squaring a vanishing coupling cannot recover it. A flat conditional mean is therefore not sufficient for the repairs to fail, whereas a vanishing physical coupling is.

\subsubsection{Normalising by a Fisher information}

A third construction perturbs the source state by an infinitesimal displacement and measures the response as a Kullback--Leibler divergence between the perturbed and unperturbed predictive densities of the target, normalised by the information-theoretic cost of the perturbation \cite{Auconi2021}. For Gaussian conditionals with state-independent variance that information response is proportional to $\left(\partial_{x_0}\E[\,y_\tau\mid x_0,y_0\,]\right)^2$, the squared derivative of the target's conditional mean with respect to the initial source state, which at short lag is the lag times the derivative of the target's drift with respect to the source. It is therefore a third functional of the physical coupling to variations of the source, and saturation suppresses it quadratically rather than linearly. The Gaussian, fixed-variance reduction is the form evaluated here; a predictive law that changes shape under the perturbation adds further terms.

\subsubsection{The common limitation}

All three repairs modify how a response to a variation of the source is weighted or expanded, but none changes what is being weighted or expanded. Where the physical coupling to variations of the source vanishes over the occupied states, as it does at saturation, the conditional source mean is flat and the integrand of Eq.~\eqref{eq:T4} vanishes pointwise, while the second-order and Fisher responses, which scale with the square of that coupling, vanish with it. A measure of sensitivity to varying the source cannot report a contribution in a state where the source, at its actual value, supplies a large forcing that does not respond to such variations. Reporting that contribution requires comparing the forcing with a stated reference level, which the attribution of Sec.~\ref{sec:adjoint} does once the reference is chosen (Sec.~\ref{sec:saturation}).

\begin{figure*}[t]
\includegraphics[width=\textwidth]{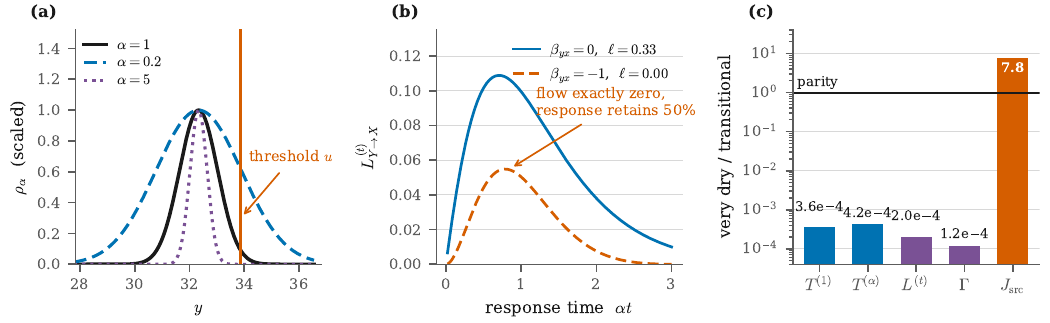}
\caption{No repair of the entropy measure reads the level. (a) Escort densities $\rho_\alpha$ of Eq.~\eqref{eq:escort} in the very dry regime of the testbed. Lowering $\alpha$ broadens the weight and raising it contracts the weight onto the mode, but the weight remains centred throughout and no $\alpha$ moves it to the threshold (orange). (b) Finite-time information response \cite{Smirnov2026,Smirnov2022} for the linear system in which the Liang--Kleeman flow $\ell$ vanishes by antisymmetric bidirectional coupling. There the repair succeeds, since at $\ell=0$ the response still retains half of its strongly coupled value. Curves are computed from the definition. The first derivative reproduces the analytic flow to $10^{-9}$, and the second-order response reproduces the published closed form to five significant figures. (c) Ranking of the very dry regime against the transitional one in the testbed, by the Liang flow $T^{(1)}$, by its R\'enyi analogue $T^{(\alpha)}$ at the best index over $\alpha\in[0.2,5]$, by the peak finite-time information response $L^{(t)}$, by the Fisher-normalised information response $\Gamma$ \cite{Auconi2021}, and by the level $\max_y|J_{\rm src}|$. The four slope-based measures agree with one another to within a factor of four and place the saturated regime between two and a half and eight thousand times below the transitional one, whereas the level places it eight times above.}
\label{fig:repairs}
\end{figure*}

The testbed of Sec.~\ref{sec:numerics} makes the size of the effect concrete. Computed on an independent ensemble of $4\times10^{5}$ soil paths, the very dry regime has $\sup_y|\partial_y j|\simeq10^{-4}$ while $j\simeq2.99$. Ranking that regime against the transitional one, $T^{(\alpha)}$ gives $3.6\times10^{-4}$ at $\alpha=1$ and at best $4.2\times10^{-4}$ over $\alpha\in[0.2,5]$, the finite-time information response gives $2.0\times10^{-4}$, and the Fisher-normalised response gives $1.2\times10^{-4}$. The level $\max_y|J_{\rm src}|$ ranks the same two regimes in the opposite order, by a factor of $7.8$ [Fig.~\ref{fig:repairs}(c)].

The claim has a qualification. Where $\partial_y j$ is appreciable but sits away from the mode, reweighting is not inert, and in the wet regime $T^{(\alpha)}$ rises by a factor of $27$ between $\alpha=5$ and $\alpha=0.2$. A repaired entropy measure thus mitigates a mismatch of location but does nothing against a mismatch of kind, in which the response has saturated, and it is the second that characterises the states in which a source supplies its largest forcing.

Two limits bound the statement. It concerns functionals of the conditional source mean, so a source acting only on the noise amplitude of the target is invisible to Eqs.~\eqref{eq:T3} and~\eqref{eq:T4}, as it is to the drift perturbation of Sec.~\ref{sec:adjoint} (Sec.~\ref{sec:scope}). And Eq.~\eqref{eq:T4} concerns a population functional, whereas the behaviour of a binned estimator of the lagged R\'enyi transfer entropy at small $\alpha$ is treated in Appendix~\ref{app:renyi}.

\subsection{Stationarity}

At stationarity $\partial_t\rho=0$, so $\partial_y J=0$ and, with the decay condition, the total current vanishes identically. The individual terms of Eq.~\eqref{eq:split} do not. A stationary state is one in which the source's transport of probability toward the threshold is exactly balanced by relaxation and diffusion, not one in which the source is inactive. This is the exact analogue of the familiar situation in which a target's entropy budget closes while its individual source contribution does not, and it is the first indication that the terms of an exact decomposition need not be causal shares.

\begin{figure}[t]
\includegraphics[width=\columnwidth]{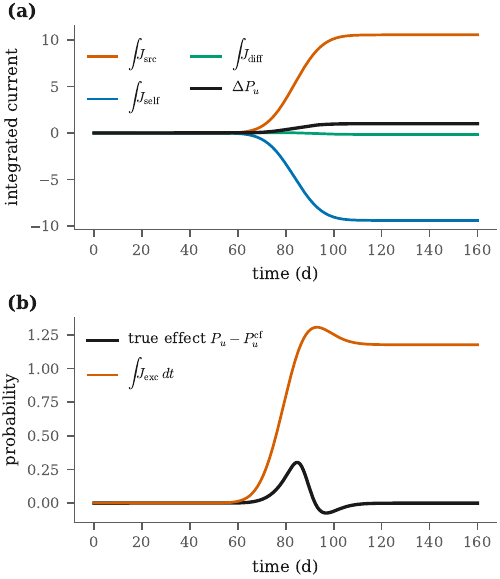}
\caption{An exact decomposition that does not attribute. (a) Time-integrated current terms at $u=29$ along the traversal. Their sum reproduces $\Delta P_u$ exactly, but the source and self terms reach $+10.57$ and $-9.41$ for a net change of $1.00$, so the terms are gross transports that nearly cancel. (b) Ground-truth counterfactual at $u=30$, in which the source response is replaced by its ensemble mean at each time. The integrated conditional-excess current reaches $1.10$ against a true peak effect of $0.30$.}
\label{fig:nonattribution}
\end{figure}

\section{An exact decomposition that does not attribute}
\label{sec:nonattr}

Integrating Eq.~\eqref{eq:T1} with the split \eqref{eq:split} gives an exact
budget for the change in exceedance probability over a finite interval,
\begin{equation}
 P_u(t_1)-P_u(t_0)=\int_{t_0}^{t_1}\!\!
 \left[J_{\rm self}+J_{\rm src}+J_{\rm diff}\right](u,t)\,dt .
 \label{eq:cumulative}
\end{equation}
It is tempting to read $\int J_{\rm src}\,dt$ as the source's contribution to the change. It is not, and the failure is not marginal. In the testbed of Sec.~\ref{sec:numerics}, a transition at $u=29$ over which $P_u$ rises by $1.000$ has $\int J_{\rm src}\,dt = 10.57$ against $\int J_{\rm self}\,dt = -9.41$, and at $u=32$ the two are $+104.0$ and $-105.2$ for a net of $0.472$ [Fig.~\ref{fig:nonattribution}(a)]. The terms exceed the quantity they sum to by one to two orders of magnitude.

The reason is that $J_{\rm src}$ is a gross transport. Where relaxation is fast, drift toward the threshold and relaxation away from it are individually large and nearly cancel, so integrating the terms separately accumulates two-way traffic rather than net effect. Equivalently, the current at time $t$ carries no information about whether the probability it moves across $u$ remains there at $t_1$, because the decomposition is local in time whereas attribution is not. The bracket structure of Eq.~\eqref{eq:split}, which records the terms of the drift a contribution came from, is not the causal structure, which records what would change if a term were altered.

The flux decomposition therefore retains its value as a diagnostic. It localises, in state space and in time, where and when a source is moving probability, and Eq.~\eqref{eq:T3} ties that diagnostic to the established entropy-transfer literature. For attribution a different object is required.

\begin{figure*}[t]
\includegraphics[width=\textwidth]{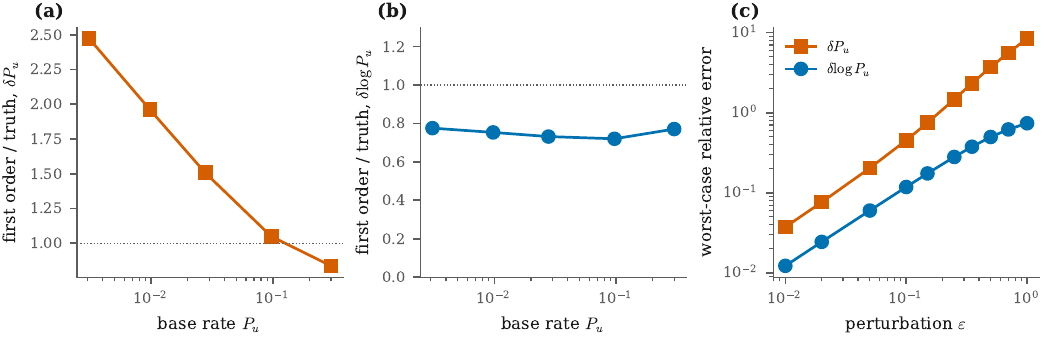}
\caption{Accuracy of the adjoint response, Eq.~\eqref{eq:response}, against the ground-truth counterfactual. (a) Read as an absolute change in $P_u$, the first-order prediction at $\varepsilon=0.25$ degrades as the event becomes rarer, overshooting the true change two-and-a-half-fold at the deepest threshold. (b) Read as a relative change in $\log P_u$, the same prediction recovers between $0.72$ and $0.78$ of the true change, with no systematic dependence on the base rate. (c) Worst-case error over the five thresholds. Convergence of both readings as $\varepsilon\to0$ verifies the identity and its implementation, and their divergence at finite $\varepsilon$ reflects the nonlinearity of the map between $P_u$ and $\log P_u$.}
\label{fig:adjoint}
\end{figure*}

\section{Attribution by adjoint response}
\label{sec:adjoint}

\subsection{The response identity}

Let $\Lgen_t$ denote the generator of Eq.~\eqref{eq:sde} and let
\begin{equation}
 \psi(\bm x,t)=\E\!\left[\chi_u(Y_{t_1})\,\middle|\,\bm U_t=\bm x\right],
 \qquad \chi_u(y)=\mathbf 1\{y>u\},
\end{equation}
so that $\psi$ solves the backward Kolmogorov equation
\begin{equation}
 \partial_t\psi+\Lgen_t\psi=0,\qquad \psi(\cdot,t_1)=\chi_u,
 \label{eq:backward}
\end{equation}
and $P_u(t_1)=\E[\psi(\bm U_{t_0},t_0)]$. Perturb the drift by
$\bm F\to\bm F+\varepsilon\,\delta\bm F$, keeping the initial law and the terminal event fixed. Writing $P_u^{\varepsilon}$ for the
exceedance probability of the perturbed process and differentiating the Feynman--Kac representation at $\varepsilon=0$ gives the standard linear-response identity \cite{GobetMunos2005}
\begin{equation}
 \boxed{\;
 \left.\frac{dP_u(t_1)}{d\varepsilon}\right|_{0}
 = \int_{t_0}^{t_1}\!\!
 \E_t\!\left[\delta\bm F(\bm U_t,t)\cdot\nabla\psi(\bm U_t,t)\right]dt \;}
 \label{eq:response}
\end{equation}
derived in Appendix~\ref{app:girsanov}. For a perturbation acting only in the target component this reduces to $\int\E_t[\delta F_Y\,\partial_y\psi]\,dt$. The contrast with Eq.~\eqref{eq:cumulative} lies in the weight. The adjoint $\partial_y\psi$ measures how much a perturbation applied at $(\bm x,t)$ propagates to the terminal event, which is precisely the information the instantaneous current discards.

Two properties matter for the intended use. Equation~\eqref{eq:response} requires only the generator and the occupied distribution, both of which a fitted drift-and-diffusion model supplies, so no counterfactual trajectory need be integrated. And because the perturbation $\delta\bm F$ is chosen, the same machinery answers different attribution questions without refitting, such as removing the state dependence of a source, replacing its forcing by a climatological level, removing the source entirely or scaling it. These are different interventions, and a stated contribution means something only together with its reference (Sec.~\ref{sec:saturation}).

\subsection{The functional in which first order is accurate}
\label{sec:functional}

Equation~\eqref{eq:response} is a first-order result, and its absolute and relative readings,
\begin{equation}
 \delta P_u \simeq \varepsilon A,
 \qquad
 \delta \log P_u \simeq \varepsilon A / P_u,
 \qquad
 A \equiv \left.\tfrac{dP_u}{d\varepsilon}\right|_0 ,
 \label{eq:twoforms}
\end{equation}
are algebraically equivalent to that order. They are not equivalent as approximations to the true finite change, because the map between $P_u$ and $\log P_u$ is nonlinear and the exact response of a tail probability to a drift perturbation is closer to exponential than to linear \cite{Ragone2018}. We find empirically that the relative form is the more uniformly accurate of the two over the perturbations tested. Over thresholds spanning two decades of exceedance probability, the error of the absolute reading grows severalfold as the event becomes rarer, whereas that of the relative reading stays nearly constant [Fig.~\ref{fig:adjoint}], as quantified in Sec.~\ref{sec:numerics}. Attribution should therefore be reported as a relative change in event probability, and the log form is carried through the remainder of the paper.

The size and sign of the remaining error follow in closed form when the perturbation acts as a shift of the target's terminal law. Let that law be Gaussian with standard deviation $s$, and let removing the source lower its mean by $\Delta>0$, which raises the standardised threshold $z=(u-\mu)/s$ to $z+\delta$ with $\delta=\Delta/s$. With $\bar\Phi$ the standard normal survival function and $h=\varphi/\bar\Phi$ its hazard rate, the first-order relative reading is $-h(z)\delta$ and the true change is $\log\bar\Phi(z+\delta)-\log\bar\Phi(z)$. Because $\log\bar\Phi$ is concave, the first-order reading understates the true change, and the fraction by which it falls short, which we call the shortfall, is
\begin{equation}
 1-\frac{h(z)\,\delta}{\log\bar\Phi(z)-\log\bar\Phi(z+\delta)}
 \;\simeq\;\frac{\delta}{2}\,\bigl[h(z)-z\bigr]
 \label{eq:meanshift}
\end{equation}
to leading order in $\delta$. Since $h(z)-z$ decreases monotonically for the Gaussian, the shortfall shrinks toward the upper tail, as $\delta/(2z)$ for large $z$, and tends to one as $P_u\to1$, where the event is nearly certain and the first-order reading becomes negligible compared with the true change. The leading-order form holds only while $\delta\,|h(z)-z|$ is small, and the exact form should be used otherwise. The same concavity gives an overstatement when a shift is added rather than removed. Neither statement extends automatically to a perturbation that changes the spread of the terminal law, which is how the perturbation of Sec.~\ref{sec:accuracy} acts.

\subsection{What the estimator requires}
\label{sec:requires}

\begin{figure*}[t]
\includegraphics[width=\textwidth]{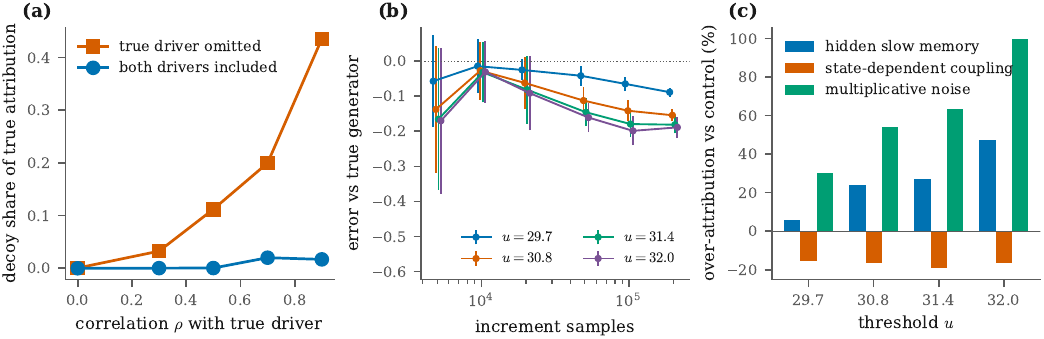}
\caption{Operating envelope. (a) An inert decoy correlated with the true driver at $\rho$ receives spurious attribution when the true driver is omitted from the drift model, and essentially none when both are included, at every correlation tested. (b) Error from estimating the generator, relative to using the true one, as mean and spread over five fits. The spread falls below five per cent above $5\times10^{4}$ increment samples, where a bias of four to twenty per cent from the fitted relaxation rate remains. (c) Over-attribution under the three misspecifications of Sec.~\ref{sec:numerics}, relative to the correctly specified control. Hidden slow memory and multiplicative noise bias the attribution upward at every threshold, and state-dependent coupling biases it downward.}
\label{fig:envelope}
\end{figure*}

Equation~\eqref{eq:response} is valid for the true generator. In application $\psi$ is built from a fitted one, and the attribution is only as good as that fit. Two conditions govern its validity, and both are properties of the observed state rather than of the attribution step. The first is completeness of the conditioning set. The identity attributes whatever perturbation direction it is given and performs no variable selection, so in the confounding structure tested in Sec.~\ref{sec:numerics} a candidate driver that is correlated with a true driver but absent from the dynamics receives a share of the attribution if the true driver is omitted from the drift model, and essentially none if it is included. The second is that the fitted generator be Markov in the observed state, because unmodelled slow memory is absorbed into the fitted relaxation rate and diffusion and biases $\psi$. Both conditions are addressed by enlarging the observed state, with the caveat, taken up in Sec.~\ref{sec:requirements}, that a variable which is itself a consequence of the source or of the target changes the estimand when it is added. Both are quantified in Sec.~\ref{sec:numerics} and Fig.~\ref{fig:envelope}.

\section{Estimation}
\label{sec:estimation}

\subsection{Fitting the generator}

Every quantity in Sec.~\ref{sec:adjoint} follows from a drift-and-diffusion model. Let $\bm X$ collect the candidate drivers, that is, the source together with any measured controls. We fit the model by regressing increments of the target on a basis in the observed state,
\begin{equation}
 \frac{\Delta Y}{\Delta t}
 = -\theta\,Y + c(\bm X) + \text{residual},
 \label{eq:driftfit}
\end{equation}
with $c(\bm X)=\sum_j c_j(x_j)$ additive across drivers and each $c_j$ expanded in a piecewise-linear (tent) basis over the observed range of that driver, and $\sigma^2$ recovered from the residual variance as $\sigma^2=\Delta t\,\Var(\text{residual})$. The tent basis is used in preference to polynomials because the response of interest is a sharp sigmoid in the driver and is poorly represented by a low-order polynomial, and in preference to a smoothing spline because it needs no smoothing penalty. Its remaining choices are the number and placement of the knots and a rule beyond the outermost knots, and they matter. In the testbed more knots lower the fitted relaxation rate (Sec.~\ref{sec:numerics}), and in the application the extrapolation rule is examined as a sensitivity (Sec.~\ref{sec:application}). Writing the drift in the form \eqref{eq:driftfit} rather than separating a relaxation target folds the intercept into $c$. The individual components $c_j$ are then defined only up to constants that can be moved between them without changing the total drift, but the perturbation field below is a difference of one component at two levels and does not depend on those constants.

The perturbation field follows from the fit. For candidate driver $j$,
\begin{equation}
 \delta F_j(\bm x,t) = \overline{c_j}(t) - c_j(x_j),
 \label{eq:perturbfield}
\end{equation}
where $c_j$ is that driver's fitted component and $\overline{c_j}(t)$ a reference level, fixed in advance and independent of $\varepsilon$. Two references are used. The ensemble mean of $c_j$ at time $t$, the reference of the testbed, removes the driver's state-dependent contribution while leaving its mean forcing intact, and so estimates the conditional-excess attribution of Eq.~\eqref{eq:T2}. The value of $c_j$ at the day-of-year climatology of $x_j$, the reference of the application in Sec.~\ref{sec:application}, removes the driver's departure from a historical baseline, including any change of its mean forcing relative to that baseline, and we call it a climatological replacement. The two differ in general, because the factual ensemble mean is not the climatology and because $c_j$ at the mean of $x_j$ is not the mean of $c_j$. In a saturated regime they differ qualitatively, the first returning nothing and the second the contribution of the saturated forcing's excess over the reference (Sec.~\ref{sec:saturation}). For the full replacement ($\varepsilon=1$) used in Sec.~\ref{sec:application}, Eq.~\eqref{eq:twoforms} gives the first-order risk ratio $P_u/P_u^{\rm cf}\simeq\exp(-A/P_u)$.

The construction corresponds to an operation already established in the information-flow literature. Replacing a driver's fitted component by a reference level is a freezing operation in the sense of Liang \cite{Liang2016}, in which the counterfactual is defined by modifying the vector field so that the driver no longer exerts its state-dependent influence, rather than by intervening on the state itself. The two constructions differ in where the freeze is placed, and the difference follows from the estimand. \citet{Liang2008} freezes the source instantaneously as a parameter, so that it retains its value but ceases to evolve, and the resulting quantity is the rate at which the source transfers information to the target, a property of the coupling as it stands. Equation~\eqref{eq:perturbfield} freezes the source's contribution at its reference level, so that the source retains its evolution but loses its anomaly. That is the operation an event attribution requires, because the question is what a realised departure from climatology cost, not how strongly the two variables are coupled.

One property carries over with the construction. Because $\delta F_j$ is built from the fitted dependence of the target's drift on
driver $j$, a driver on which that drift does not depend contributes
$\delta F_j\equiv0$ identically, and by Eq.~\eqref{eq:response} the attribution
is then exactly zero. This is the principle of nil causality
\cite{Liang2016,LiangKleeman2005}, obtained here without a separate argument
(in a fitted model the attribution is near zero rather than zero, as
Sec.~\ref{sec:numerics} confirms). It
is a property of the perturbation field rather than of the functional being
attributed, so it holds for the exceedance probability of Eq.~\eqref{eq:T1}
exactly as it does for the marginal entropy rate, and for functionals of the
path such as an occupation time, a first-passage probability or a committor once
the backward problem is modified accordingly (a running cost for the first,
absorbing boundaries for the others). \citet{Liang2016}
establishes the property separately for deterministic mappings, continuous
flows and stochastic systems, whereas in the response formulation it is
structural and the generality is inherited rather than proved again. Like
Liang's, the statement concerns the direct pathway. A driver that acts on the
target only through another state variable, or only through the diffusion,
contributes $\delta F_j\equiv0$ to the target's drift and receives no attribution
under this perturbation, however strong its total influence.

\subsection{Computing the adjoint}

Two routes are available. The general one integrates
Eq.~\eqref{eq:backward} backward on a grid, with one practical caution.
First-order upwind advection injects numerical diffusion of order
$|\bar F|\,\Delta y/2$, which in our testbed exceeds the physical diffusion by nearly
an order of magnitude and produces errors that do not vanish as the perturbation
does. Refining the grid reduces this error only in proportion to the grid
spacing, and because $\chi_u$ is a step, $\partial_y\psi$ approaches a delta
function as $t\to t_1$, so convergence is slow and expensive. A flux-limited or
semi-Lagrangian scheme, with the terminal step integrated over each cell, is the
practical choice if this route is taken.

The second route is available whenever the target is conditionally linear given
the remaining coordinates, which covers the common case of an
Ornstein--Uhlenbeck target driven by a nonlinear function of a source that
evolves without feedback from the target and whose noise is independent of the
target's. Then the
terminal probability is available in closed form: conditional on a driver path,
\begin{align}
 Y_{t_1} &= a(t)\,Y_t + I(t) + \text{noise},\qquad
 a(t)=e^{-\theta(t_1-t)},\nonumber\\
 I(t) &= \int_t^{t_1}\! e^{-\theta(t_1-s)}\,c(\bm X_s)\,ds,\qquad
 s^2(t)=\frac{\sigma^2\!\left(1-a^2\right)}{2\theta},
\end{align}
so that, with $\Phi$ and $\varphi$ the standard normal distribution and density
and $\E_{\bm X}$ the expectation over driver paths on $[t,t_1]$ conditional on the
state at $t$,
\begin{equation}
 \psi = \E_{\bm X}\!\left[\Phi\!\left(\tfrac{M-u}{s}\right)\right],
 \quad
 \partial_y\psi = \E_{\bm X}\!\left[\varphi\!\left(\tfrac{M-u}{s}\right)
 \frac{a}{s}\right],
 \label{eq:semianalytic}
\end{equation}
with $M(y,t)=a(t)y+I(t)$. This is exact in the linear direction and Monte Carlo
in the nonlinear one, mirroring the representation used for the forward density,
and no numerical diffusion is introduced. Consistency with the forward
calculation is not automatic, however. When the model is integrated on a discrete
step, as in any application, $a$, $I$ and $s^2$ must be evaluated with the same
discrete kernel, so that $\psi$ is the backward solution of the model actually
simulated. Pairing the exponential kernel above with an Euler forward step can shift
$M$ by a few tenths of a degree in the hourly application of
Sec.~\ref{sec:application}, which is enough to bias the response at the tail
thresholds by tens of per cent.

\subsection{Uncertainty}

Writing Eq.~\eqref{eq:response} as an average over trajectories gives its
Monte Carlo error directly. Accumulating a per-trajectory
contribution
\begin{equation}
 C_i = \int_{t_0}^{t_1}\!\delta F(\bm U^{(i)}_t,t)\,
 \partial_y\psi(\bm U^{(i)}_t,t)\,dt
\end{equation}
gives $A=\overline{C_i}$ with Monte Carlo standard error
$\mathrm{sd}(C_i)/\sqrt{N}$. The relative reading $A/P_u$ is formed from the same
trajectories, and its standard error follows from the influence variable
$(C_i-A)/P_u-A(Q_i-P_u)/P_u^2$, with $Q_i$ the conditional exceedance probability
of trajectory $i$. Trajectories simulated from one fitted generator are
independent, so this error needs no bootstrap or block structure, and
Sec.~\ref{sec:numerics} verifies that the standard error of $A$ reproduces the
spread across independent ensembles. It does not include the uncertainty of the fit itself, which all
trajectories share, and which Sec.~\ref{sec:application} estimates by refitting on
resampled years.

Two further requirements are specific to a tail estimand. The threshold must be
fixed, either physically or as a quantile of a fixed baseline, because a
threshold re-estimated at each time defines a moving population and the
identity \eqref{eq:T1} no longer refers to a single event. And the estimator should not
inherit support screening designed for a bulk quantity. A rule that withholds
periods whose occupancy departs from a reference will preferentially discard the
anomalous periods, which for a tail estimand removes the data the analysis
exists to use. This is different from a lack of support. Where the event, or its
counterfactual, visits combinations of the drivers that are absent from the
training data, the fitted components are extrapolated and the attribution is
conditional on the extrapolation rule. That dependence should be reported, not
screened away (Sec.~\ref{sec:application}).

\section{Numerical validation}
\label{sec:numerics}

\subsection{Testbed}

We use a manufactured soil-moisture--temperature system with a saturating
coupling, chosen because its conditional structure permits every quantity to be
evaluated from the known dynamics rather than from a fitted regression. With
$S=\operatorname{logistic}(U)$,
\begin{align}
 dU &= -\theta_U\!\left[U-\nu(t)\right]dt + \sigma_U\,dW_U,\nonumber\\
 dT &= \left[-\theta_T(T-T_{\rm ref}) + g(S)\right]dt + \sigma_T\,dW_T,
 \label{eq:testbed}\\
 g(S) &= \frac{3}{1+\exp\!\left[(S-0.35)/0.055\right]},\nonumber
\end{align}
with time in days and temperature in $^{\circ}$C, $\theta_U=0.08$~d$^{-1}$,
$\theta_T=0.25$~d$^{-1}$, $T_{\rm ref}=20\,^{\circ}$C,
$\sigma_T=\sqrt{0.5}\times0.7\,^{\circ}$C\,d$^{-1/2}$, $g$ in $^{\circ}$C\,d$^{-1}$,
and $\sigma_U$ set so that $U$ has stationary standard deviation $0.35$. The soil
equilibrium $\nu(t)$ is held at $\operatorname{logit}(0.74)$,
$\operatorname{logit}(0.35)$ and $\operatorname{logit}(0.04)$ for the stationary
wet, transitional and very dry regimes, and traversed linearly from
$\operatorname{logit}(0.74)$ to $\operatorname{logit}(0.04)$ between days $10$
and $130$ for the nonstationary experiments. The adjoint and operating-envelope experiments use a
five-day window, comparable to the lead times of Sec.~\ref{sec:application},
starting from the stationary transitional regime, and a perturbation
$\varepsilon=0.25$ of Eq.~\eqref{eq:perturbfield} with the ensemble-mean
reference. This is a consistency model rather than a simulation of a
physical heatwave, and its role is to supply a system in which the
counterfactual is computable exactly.

Conditional on a soil path, $T$ is Gaussian, so the target density is represented
exactly as a Gaussian mixture over an ensemble of soil paths and no density
estimation enters the verification. Ensembles have $10^{5}$ to $4\times10^{5}$
paths, with the random seeds fixed in the archived scripts.

\subsection{The identities}

Along a nonstationary wet-to-dry traversal, $dP_u/dt=J(u,t)$ holds to a median
relative error of $1.9\times10^{-4}$ and a maximum of $2.6\times10^{-3}$ over
five thresholds and seventeen times, the residual being the finite-difference
floor of the reference derivative [Fig.~\ref{fig:decomposition}(c)]. The bridge
identity \eqref{eq:T3} holds to relative errors of $3\times10^{-11}$,
$1\times10^{-14}$ and $4\times10^{-7}$ in the wet, transitional and very dry
regimes, and so nonlinearly rather than only in a Gaussian limit. A source with
nonzero mean but independent of the target gives a conditional-excess current
that vanishes to machine precision while its mean-forcing current does not,
confirming Eq.~\eqref{eq:T2}.

At stationarity the total current vanishes to $9.5\times10^{-6}$,
$1.7\times10^{-4}$ and $1.1\times10^{-6}$ in the three regimes while the source
current reaches $0.0036$, $0.2233$ and $1.7029$ [Fig.~\ref{fig:decomposition}(a)].
The dry regime, in which the entropy transfer is smallest, carries the largest
source current of the three.

\subsection{The decomposition does not attribute}

The cumulative budget \eqref{eq:cumulative} closes to between $4\times10^{-6}$
and $1.2\times10^{-5}$, while its terms show the gross transports reported in
Sec.~\ref{sec:nonattr} [Fig.~\ref{fig:nonattribution}(a)]. Against a ground-truth
counterfactual in which the source response is replaced by its ensemble mean at
each time, the conditional-excess current integrated up to the peak of the true
effect reaches $1.10$ against a true peak effect of $0.30$ at $u=30$, and $0.45$
against $0.085$ at $u=32$ [Fig.~\ref{fig:nonattribution}(b)]. It overshoots by a
factor of four to five, and its time course bears no resemblance to that of the
effect.

\subsection{Accuracy of the adjoint}
\label{sec:accuracy}

At small perturbation the adjoint prediction matches the counterfactual to a
median of $1.2\%$ and a maximum of $3.7\%$ across five thresholds, with base rates from $0.3$ to $0.003$, which verifies
Eq.~\eqref{eq:response} and its implementation. At finite perturbation the
absolute and relative readings separate, and the worst error over the five
thresholds grows to $1.5$ for $\delta P_u$ but remains at $0.28$ for
$\delta\log P_u$ at $\varepsilon=0.25$, and to $8.4$ against $0.74$ at
$\varepsilon=1$ [Fig.~\ref{fig:adjoint}]. Across six independent ensembles the
relative reading gives $0.771$, $0.718$, $0.730$, $0.753$ and $0.776$ of the true
change at $\varepsilon=0.25$, each $\pm0.001$. The shortfall, one minus that
fraction, is twenty-two to twenty-eight per cent across the five thresholds. It is
therefore a reproducible first-order truncation rather than sampling error, and
can be reported as a characterised bias for this perturbation. The perturbation
moves each path's forcing toward the ensemble mean and leaves the mean forcing
unchanged, so it narrows the terminal law rather than shifting it, and
Eq.~\eqref{eq:meanshift} does not apply to it. The same ensembles validate the analytic
standard error, the ratio of the spread across replicates to the mean
per-trajectory standard error being $1.07$, $1.00$, $0.92$, $1.01$ and $0.79$
at the five thresholds.

\subsection{Operating envelope}

Table~\ref{tab:envelope} collects the conditions under which the estimator is
usable, each established against the same ground-truth counterfactual.

Two results deserve comment. First, an inert candidate driver correlated with
the true driver receives up to $0.44$ of the true attribution when the true
driver is omitted from Eq.~\eqref{eq:driftfit}, and at most $0.02$ when both are
included, at every correlation up to $0.9$ [Fig.~\ref{fig:envelope}(a)]. The
method therefore fails under \emph{omitted} confounding rather than under
confounding as such, and the remedy is the complete conditioning set required
in Sec.~\ref{sec:requires}.

Second, the three misspecifications do not act alike
[Fig.~\ref{fig:envelope}(c)]. We simulate hidden slow memory as an unobserved
Ornstein--Uhlenbeck forcing of the target with a fifty-day timescale and a
standard deviation of $0.45\,^{\circ}$C\,d$^{-1}$, state-dependent coupling as
$g(S)\,[1+0.04\,(T-26)]$, and multiplicative noise as a noise amplitude
$\sigma_T[1+0.8\,(g(S)/1.5-1)]$ that scales with the source. Relative to the
correctly specified control, hidden slow memory biases the attribution
\emph{upward} by six to forty-eight per cent and multiplicative noise by thirty
to one hundred per cent, the opposite direction to the first-order truncation,
whereas state-dependent coupling biases it downward by fifteen to nineteen per
cent. These directions belong to the constructions tested. They depend on the
signs of the couplings and correlations and are not general, so a source that
enters multiplicatively should be carried in the fitted model rather than
corrected for afterwards. Unmodelled slow memory is the most dangerous case, not
because its error is the largest but because it is the hardest to see. It leaves
the squared-residual diagnostic untouched, and its trace in the residuals is a
weak but persistent autocorrelation, between $0.01$ and $0.03$ at lags of one to five days.
A portmanteau statistic over those lags detects it at $10^{5}$ increments
($Q=1373$, against $37$ to $51$ in the other three worlds and a one-per-cent
critical value near $65$), but a visual check of the residuals would not. Its
other trace is a fitted relaxation rate of $0.111$ against a true $0.25$, so we
recommend that the fitted relaxation timescale and the residual autocorrelation
always be reported and checked against physical expectation. Multiplicative
noise, by contrast, shows in the correlation between squared residuals and state
($-0.363$, against less than $0.01$ in magnitude for every other case), although
that statistic tests only one linear form of state-dependent variance.

Estimating the generator rather than using the true one lowers the attribution
by four to twenty per cent above $5\times10^{4}$ increment samples, a bias that
comes mostly from the fitted relaxation rate ($0.236$ against $0.25$) and does not
shrink with more data, while the spread across fits falls below five per cent.
At $5\times10^{3}$ samples the spread grows to about twenty per cent
[Fig.~\ref{fig:envelope}(b)]. Only a small part of the relaxation bias comes from
the sampling step, since an increment regression at step $\Delta t=0.1$~d
estimates $(1-e^{-\theta\Delta t})/\Delta t=0.247$ rather than $0.25$. On three
further training sets of the same size, regressing on the true source function
instead of the tent basis recovers $0.243$, whereas the tent basis fitted to the
same sets gives $0.229$ with fourteen knots, $0.220$ with twenty-eight and $0.203$
with fifty-six (each $\pm0.002$ to $\pm0.005$ across sets), so most of the bias
comes from representing the source by a flexible basis and grows with its
flexibility.

\begin{table}[t]
\caption{Operating envelope of the adjoint attribution, established against a
ground-truth counterfactual in the testbed \eqref{eq:testbed}.}
\label{tab:envelope}
\begin{tabular}{@{}p{0.58\columnwidth}l@{}}
\toprule
Condition & Effect \\
\midrule
Correct model, $>5\times10^{4}$ samples
 & $\sim$35\% under\footnotemark[1] \\
$5\times10^{3}$ increment samples
 & $\pm$20\% spread \\
Correlated driver omitted ($\rho=0.9$)
 & up to 0.44 spurious \\
Both drivers in the conditioning set
 & $\le0.02$ spurious \\
Hidden slow memory
 & $+6$ to $+48$\% \\
State-dependent coupling
 & $-19$ to $-15$\% \\
Multiplicative noise
 & $+30$ to $+100$\% \\
\bottomrule
\end{tabular}
\footnotetext[1]{At the thresholds tested here, combining the first-order
shortfall of twenty-two to twenty-eight per cent with the bias from the fitted
relaxation rate. Misspecification effects are relative to this control.}
\end{table}

\subsection{The saturated regime and the reference}
\label{sec:saturation}

The regime that motivated the paper is the one in which the choice of reference
matters most. Table~\ref{tab:saturation} repeats the five-day experiment of
Sec.~\ref{sec:accuracy} from the very dry and the transitional regimes with three
perturbations of the source forcing: the ensemble-mean reference used so far, a
fixed climatological reference equal to the mean forcing of the transitional
regime ($1.48\,^{\circ}$C\,d$^{-1}$), and full removal of the source. The exact
counterfactual is again available because the target is conditionally Gaussian.

With the ensemble-mean reference, the attribution in the very dry regime is of
order $10^{-3}$ per unit $\varepsilon$, as small as the entropy flow and for the
same reason, since a saturated forcing has almost no anomaly about its mean. With
the climatological reference the same regime returns a first-order response of
$-7$ to $-19$ per unit $\varepsilon$ at base rates from $0.3$ to $0.003$, and
replacing a quarter of the forcing's departure from that reference lowers the
exceedance probability by a factor of $15$ to $340$. The contribution that the
entropy flow cannot see is therefore visible to the adjoint response once the
question states what the forcing is compared with, and the conditional-excess
reference, the natural one for a coupling diagnostic, is blind in exactly the
regime that motivated this paper.

Because the saturated forcing barely varies across paths, the fixed-reference
perturbation is a pure shift of the terminal law, and its shortfall, falling from
$35$ to $19$ per cent toward the tail at $\varepsilon=0.25$ and from $70$ to $49$
per cent at $\varepsilon=1$, agrees with the exact form of Eq.~\eqref{eq:meanshift} to the precision shown.
Two limits follow. For a large intervention the first-order reading has the right
sign but falls short by a factor of two to three for this reference, and by more
for full removal, so the direct counterfactual should be computed. And where the event is nearly certain the first-order reading
fails altogether. At $u=29\,^{\circ}$C, which the very dry regime exceeds with
probability $0.99999$, the first-order response to the climatological replacement
is $-3\times10^{-4}$ per unit $\varepsilon$, whereas the full replacement lowers
$\log P_u$ by $3.7$.

\begin{table*}[t]
\caption{First-order and exact responses of $\log P_u$ to three perturbations of
the source forcing over the five-day window, from the very dry and transitional
regimes of the testbed. Ranges run over five thresholds with base rates from $0.3$
to $0.003$, given in that order where the dependence is monotone and as an
interval (with a dash) where it is not. The shortfall is one minus the ratio of the first-order
to the exact change in $\log P_u$, and a negative shortfall means that the first-order
reading overstates the change. In the transitional regime the climatological
reference is that regime's own mean forcing and coincides with the ensemble-mean
row.}
\label{tab:saturation}
\begin{tabular}{@{}llcccc@{}}
\toprule
Regime & Reference & $\partial_\varepsilon\log P_u$ & $\delta\log P_u$, $\varepsilon=0.25$
 & Shortfall, $\varepsilon=0.25$ & Shortfall, $\varepsilon=1$ \\
\midrule
Very dry, $T^{(1)}=7.7\times10^{-5}$ & ensemble mean & $-10^{-4}$ to $-2\times10^{-3}$
 & $>-5\times10^{-4}$ & $-11$\% & $-68$ to $-64$\% \\
 & climatological & $-7.1$ to $-18.8$ & $-2.7$ to $-5.8$ & $35$ to $19$\% & $70$ to $49$\% \\
 & removal & $-14.1$ to $-37.2$ & $-7.4$ to $-13.7$ & $52$ to $32$\% & $83$ to $66$\% \\
Transitional, $T^{(1)}=0.23$ & ensemble mean & $-0.51$ to $-9.5$ & $-0.16$ to $-3.06$
 & $22$--$28$\% & $74$ to $56$\% \\
 & removal & $-2.1$ to $-20.5$ & $-0.87$ to $-8.2$ & $38$--$48$\% & $92$ to $74$\% \\
\bottomrule
\end{tabular}
\end{table*}

\section{Discussion}

Two objects have been separated here that are easily conflated. The
probability current at a threshold, decomposed by the terms of the drift, is an
exact and local diagnostic of where and when a source moves probability across
that threshold. The adjoint response, a first-order quantity that is global in
time, says what would change if the source were altered, and it is the object
that answers an attribution question. That treating the first as the second
fails by more than an order of magnitude even in an exactly solvable system
(Sec.~\ref{sec:nonattr}) is a general caution that an exact decomposition of a
rate is not a decomposition of the level it accumulates to.

\begin{figure*}[t]
\includegraphics[width=\textwidth]{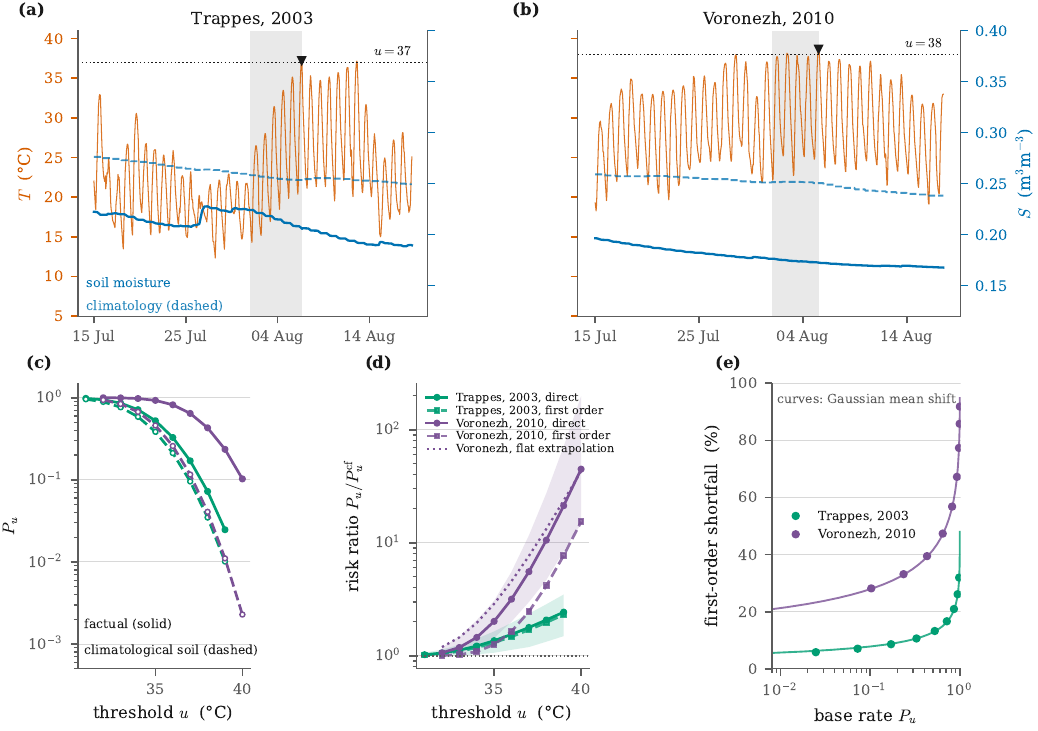}
\caption{Attribution in two observed heatwaves. (a),(b) Hourly 2-m
temperature (orange) and root-zone soil moisture (blue, solid) at the two grid
points, with the day-of-year soil-moisture climatology of the baseline summers,
event year excluded (blue, dashed), and the dotted line marks one of the
thresholds used. The shaded band runs from $t_0$ to $t_1$, the latter marked by a
triangle and fixed at the hour of the observed temperature maximum. Soil moisture
is below climatology throughout and draws down across the window. (c) Exceedance
probability at $t_1$ against threshold in the factual ensemble, whose soil
moisture is simulated under the fitted model (solid), and with the soil forcing
replaced by its climatological level (open, dashed). On the logarithmic axis the
vertical separation of each pair is the logarithm of the ratio plotted in (d).
(d) Risk ratio against threshold from the direct counterfactual (solid, circles),
with its 90\% year-block bootstrap range shaded, and from the first-order adjoint
(dashed, squares). The dotted line repeats the 2010 direct counterfactual with each
fitted component held constant beyond its outermost knot instead of continued
linearly. (e) First-order shortfall, $1-\log RR_{\rm adjoint}/\log RR_{\rm
direct}$, against base rate, with the exact form of the Gaussian mean-shift
formula, Eq.~\eqref{eq:meanshift}, for each site (curves).}
\label{fig:application}
\end{figure*}

\subsection{Application to two observed heatwaves}
\label{sec:application}

This section applies the estimator to observations, to two events for which the
land-surface contribution has been argued on physical grounds
\cite{Fischer2007,Miralles2014}, in line with observational evidence that dry
soils amplify hot extremes \cite{Hirschi2011}. The two events are the western
European heatwave of August 2003 and the western Russian heatwave of August 2010.
No ground-truth counterfactual exists for either, so nothing here validates the
attribution against a real intervention. What the application does is exercise
the estimator on data with the memory, forcing and confounding the testbed lacks,
and expose a threshold dependence that the synthetic experiments, run at fixed
thresholds, could not.

\subsubsection{Data and construction}

The target $T$ is hourly ERA5-Land \cite{MunozSabater2021} $2$-m temperature and
the source $S$ is root-zone soil moisture, the thickness-weighted mean (weights
$0.07$, $0.21$ and $0.72$) of volumetric soil water in layers $1$--$3$ over
$0$--$100$~cm. ERA5-Land is an offline replay of the ECMWF land-surface model
driven by ERA5 atmospheric forcing \cite{MunozSabater2021}. Its $2$-m temperature
is diagnosed between the prescribed lowest atmospheric level and the simulated
land surface, so the land state can influence it locally, but no feedback of the
land on the atmosphere above is simulated. Two exogenous controls taken from ERA5
\cite{Hersbach2020} enter the conditioning set, the $500$-hPa geopotential height
$Z$ (in metres), a circulation index, and the $850$-hPa temperature $H$, a proxy
for the multi-day boundary-layer heat accumulation that Sec.~\ref{sec:numerics}
identifies as the hidden-slow-memory failure mode. Both carry the land--atmosphere
coupling of the ERA5 system itself. A soil-to-temperature relation fitted across
the two products therefore mixes the local land influence represented in
ERA5-Land with forcing common to both, and the exercise is a conditional
attribution within a fitted, reanalysis-based model rather than an identified
response of the atmosphere to the soil. The series are taken at the nearest point
of the analysis-ready grids ($0.1^{\circ}$ for ERA5-Land, $0.25^{\circ}$ for ERA5)
to $48.8^{\circ}$N $2.0^{\circ}$E for 2003 and $51.7^{\circ}$N $39.2^{\circ}$E for
2010, hourly in UTC over June--August of $1979$--$2022$, with $Z$ and $H$
six-hourly and interpolated linearly to the hour. This gives $9.7\times10^{4}$
hourly samples per site, of which $9.5\times10^{4}$ increments between
consecutive hours of the same summer enter the fit.

The generator is fitted as in Sec.~\ref{sec:estimation} on the baseline summers
with the event year excluded, so the event does not inform the model used to
attribute it. Three features are added because the target is hourly. The diurnal
cycle enters through two harmonics of the hour of day, in an additive term, in the
relaxation rate and in the soil and $850$-hPa components, which take the form
$c_j(x_j,t)=\sum_l B_l(x_j)\,[v_{l0}+\sum_{k=1,2}(a_{lk}\sin k\omega t+b_{lk}\cos
k\omega t)]$ with $B_l$ the tent basis and $\omega=2\pi$~d$^{-1}$, so that dry soil
can force temperature more strongly in the afternoon than at night. The drift
remains additive across drivers, and the perturbation field is still a difference
of one component at two levels. Two harmonics of the day of year carry the
seasonal cycle. And the temperature residuals are not white. Their autocorrelation
is $0.47$ and $0.48$ at one hour and remains between $0.1$ and $0.25$ at multiples
of a day,
so the noise is given an amplitude that depends on the hour of day and the
empirical autocorrelation of the standardised residuals. Temperature then remains
conditionally Gaussian given a soil path, and $s^2(t)$ in
Eq.~\eqref{eq:semianalytic} is computed from that autocorrelation, which is
equivalent to carrying the residual memory in auxiliary linear states. The soil
equation is fitted in the same way, with temperature and $Z$ as drivers and no
precipitation input. Each component uses fourteen knots evenly spaced between the
$0.5$th and $99.5$th percentiles of its driver and is continued linearly beyond
the outermost knots, with the alternative of holding it constant examined below.
The model is integrated on the hourly step on which it was fitted, and $\psi$ is
evaluated with the same discrete kernel.

The estimand requires an initial and a terminal time. We take $t_0$ as $00$:$00$
UTC on 1 August of the event year and $t_1$ as the hour at which the observed
temperature attained its maximum over 1--15 August, which is 6 August 2003 at
$15$:$00$ and 5 August 2010 at $12$:$00$, giving lead times of $5.6$ and $4.5$
days. The attributed quantity is
\begin{equation}
 P_u=\Pr\bigl[T(t_1)>u\;\big|\;T(t_0),\,S(t_0);\;Z,\,H\text{ on }[t_0,t_1]\bigr],
 \label{eq:appestimand}
\end{equation}
with the circulation and the $850$-hPa temperature held at their observed values
in both the factual and the counterfactual ensemble. The terminal hour is selected
retrospectively and then held fixed, so $P_u$ is the probability of exceeding $u$
at that hour, not the probability that the maximum falls there, and not a
frequency over a season. The exercise is a retrospective, atmosphere-conditioned
comparison of two model states.

The counterfactual replaces the fitted soil forcing $c_S(S_t,t)$ along each path by
$c_S$ evaluated at the day-of-year soil-moisture climatology of the baseline
summers, a reference fixed in advance and independent of $\varepsilon$, and so is
the climatological replacement of Sec.~\ref{sec:estimation}. That $H$ is held at
its observed, heatwave-contaminated values is deliberate and is revisited below.
Soil moisture is simulated forward under the fitted model rather than prescribed,
so that factual and counterfactual ensembles obey the same dynamics. Temperature
feeds back on soil moisture through the soil equation, which
Eq.~\eqref{eq:semianalytic} neglects, but over these windows the counterfactual
soil paths differ from the factual ones by at most $1.3\times10^{-3}$~m$^3$m$^{-3}$
and no exceedance count changes when the soil paths are held fixed. The
probabilities below are therefore evaluated by integrating the temperature noise
exactly along $2\times10^{5}$ simulated soil paths, so that the direct
counterfactual and the first-order adjoint differ only by the truncation of the
latter. A brute-force simulation of the coupled model agrees with them within its
Monte Carlo error.

\subsubsection{The fitted generator}

The fitted relaxation timescales, averaged over the day, are $\tau_T=0.76$~d and
$\tau_S=187$~d at the 2003 site and $\tau_T=0.57$~d and $\tau_S=108$~d at the 2010
site, with $R^{2}=0.73$ and $0.72$ for the temperature drift. Refitting on
resampled years gives $90\%$ ranges of $0.72$--$0.78$~d and $131$--$319$~d at the
first site and $0.55$--$0.59$~d and $75$--$159$~d at the second. Checked against
physical expectation, as Sec.~\ref{sec:numerics} recommends, both temperature
timescales are of order a day and both soil timescales of order months. We read
nothing into the difference in $\tau_S$, which is recovered from hourly increments
of a slowly varying, strongly seasonal quantity and is poorly identified. The
heteroscedasticity diagnostic is $+0.006$ and $-0.029$.

Leave-one-year-out hindcasts test the fitted model in the conditional setting of
Eq.~\eqref{eq:appestimand}. For each baseline year the model is refitted without
that year and run from six summer dates to the lead time and hour of the event.
At the 2003 site the observed temperature falls inside the central $90\%$
predictive interval in $88\%$ of $258$ hindcasts and inside the central $50\%$ in
$55\%$, with a mean bias of $+0.14\,^{\circ}$C and a ratio of root-mean-square
error to predictive spread of $1.10$. At the 2010 site the figures are $91\%$,
$60\%$, $+0.09\,^{\circ}$C and $0.93$. Without the diurnal modulation and the
residual autocorrelation the predictive law is too narrow ($90\%$ coverage of
$75\%$ and $84\%$, ratios of $1.43$ and $1.20$), and the observed peak lies $3.7$
and $2.5$ predictive standard deviations above the model's mean, against $1.1$ and
$0.5$ with them. The soil ensemble is too narrow at five days ($90\%$ coverage of
$60\%$ and $51\%$). Rain, which the soil equation does not represent, raised soil
moisture in $13\%$ of the hindcast windows, but coverage is still only $66\%$ and
$58\%$ without them, so the soil noise estimated from hourly increments also
understates the variability of multi-day drying. Soil moisture shows no rise
during either event window, and the sensitivity of the attribution to the
simulated soil path is examined below by prescribing the observed one.

Figure~\ref{fig:application}(a,b) shows the two events. Soil moisture at $t_0$ is
below the day-of-year climatology by $0.034$~m$^3$m$^{-3}$ in 2003 and by
$0.075$~m$^3$m$^{-3}$ in 2010, $0.75$ and $1.84$ standard deviations of its
baseline distribution on that date. The two events differ in how far they lie from
the training data. At the 2003 site the simulated soil states stay within the
knots of $c_S$, and the $850$-hPa temperature exceeds the outermost knot of $c_H$
in $31\%$ of the event's hours. At the 2010 site every simulated soil state lies
below the driest knot ($0.180$~m$^3$m$^{-3}$), and the observed soil moisture fell
to $0.172$~m$^3$m$^{-3}$, below anything in the baseline summers. The $850$-hPa
temperature exceeds the outermost knot throughout the window and the baseline
maximum at its peak, and the peak $2$-m temperature, $38.7\,^{\circ}$C, is
$2.0\,^{\circ}$C above the warmest hour of the baseline summers. The 2010
attribution is therefore an extrapolation of the fitted components, and it is
reported together with the alternative rule of holding each component constant
beyond its outermost knot.

\subsubsection{Attribution and its threshold dependence}

Figure~\ref{fig:application}(c) gives the two probabilities and
Fig.~\ref{fig:application}(d) the attributed risk ratio $P_u/P_u^{\rm cf}$ against
threshold. The deficit warms the peak hour by $0.70\,^{\circ}$C on average at the
2003 site ($90\%$ bootstrap range $0.44$--$0.91\,^{\circ}$C) and by
$2.9\,^{\circ}$C at the 2010 site ($2.0$--$4.1\,^{\circ}$C, and $2.2\,^{\circ}$C
under the flat rule). At the 2003 site the direct counterfactual rises from $1.02$
at $u=31\,^{\circ}$C to $1.78$ at $37\,^{\circ}$C, the observed peak, and $2.4$ at
$39\,^{\circ}$C, while the base rate falls from $0.98$ to $0.17$ and $0.025$. The
bootstrap range at $37\,^{\circ}$C is $1.20$--$2.36$, and holding the components
constant beyond their knots gives $1.73$. At the 2010 site the ratio rises from
$1.07$ at $32\,^{\circ}$C to $10.5$ at $38\,^{\circ}$C, just below the observed
peak of $38.7\,^{\circ}$C (bootstrap range $4.7$--$27$), and $45$ at $40\,^{\circ}$C, with $13$ and $43$ under the flat rule,
while the base rate falls from $0.999$ to $0.43$ and $0.10$. The fitted soil
equation also dries the 2010 soil faster than observed, and prescribing the
observed soil path lowers the ratio at $38\,^{\circ}$C to $8.6$, whereas at the
2003 site it changes the ratio at $37\,^{\circ}$C by $2\%$. The 2010 numbers carry
the extrapolation described above and should be read as an order of magnitude.

The ratio alone conceals the magnitudes. At the 2003 site the observed peak of
$37\,^{\circ}$C carries $P_u=0.17$ against a counterfactual $0.096$, and at the
2010 site $38\,^{\circ}$C carries $0.43$ against $0.041$. The soil-moisture
deficit is the same in every one of these calculations, and what changes is the
probability it is asked to move.

The mechanism is elementary and worth stating because it governs how such a
number should be quoted. The deficit acts as a nearly uniform shift of the
terminal law, since the spread of the integrated perturbation across soil paths is
small compared with its mean. A fixed shift of the mean produces a relative change
in exceedance probability that grows as the threshold moves into a thinning tail
\cite{Sardeshmukh2015}. The ratio then increases with $u$ for a Gaussian terminal
law, or for any law with an increasing hazard rate, which is a sufficient
condition and not a general property, since a mixture of well-separated components
can violate it. Here the Gaussian description is quantitative. A Gaussian with the
ensemble's mean and spread ($35.1$ and $1.97\,^{\circ}$C in 2003, $37.7$ and
$1.83\,^{\circ}$C in 2010), shifted by the mean warming, reproduces every ratio in
Fig.~\ref{fig:application}(d) to within $3\%$ and every shortfall in
Fig.~\ref{fig:application}(e) to within one percentage point. The threshold
dependence is therefore expected even in the simplest mean-shift model and is not
evidence of a further soil-moisture mechanism. Conversely, the mean warming at the
peak hour and the predictive spread form a threshold-free summary from which the
ratio at any threshold follows. A soil-moisture contribution reported as a ratio
without its threshold is underspecified, whereas here, where the shift is nearly
uniform and the terminal law nearly Gaussian, one reported as a shift together
with its spread is not.

\subsubsection{Behaviour of the first-order estimate}

Both curves in Fig.~\ref{fig:application}(d) are shown twice, once from the
adjoint of Sec.~\ref{sec:adjoint} and once from the direct counterfactual under the
same fitted generator. The adjoint lies below the direct estimate at every
threshold, the sign expected for removing a shift from a log-concave law
(Sec.~\ref{sec:functional}). Figure~\ref{fig:application}(e) plots the shortfall
against base rate. At the 2003 site it falls from $32\%$ at $31\,^{\circ}$C, where
the effect is a few per cent and the event nearly certain, to $9\%$ at
$37\,^{\circ}$C and $6\%$ at $39\,^{\circ}$C. At the 2010 site, where the shift is
four times larger relative to the predictive spread, it falls from $92\%$ at
$32\,^{\circ}$C to $39\%$ at $38\,^{\circ}$C and $28\%$ at $40\,^{\circ}$C. Both
sequences follow the exact form of Eq.~\eqref{eq:meanshift}. The first-order reading is thus most
accurate in the upper tail, where attribution questions are usually asked, and
least accurate where the event is nearly certain or the shift is large. The
pattern differs from the testbed of Sec.~\ref{sec:accuracy}, where the shortfall
was nearly independent of base rate, and the reason is the geometry of the
perturbation, which there narrows the terminal law and here shifts it. Where the
direct counterfactual is affordable it should be computed. Where it is not, the
first-order number should be reported with its base rate and the standardised
shift, from which the exact form of Eq.~\eqref{eq:meanshift} gives the expected
truncation.

\subsubsection{What this does and does not establish}

Four limits bound what has been shown. First, the adjoint and the direct
counterfactual share the fitted generator, so their agreement tests the
first-order truncation of that generator, with the feedback of temperature on soil
moisture shown separately to be negligible, and not whether the generator
represents the system. The hindcasts test the latter only in the sense of
calibration.

Second, the conditioning set $(T,S,Z,H)$ is certainly incomplete, since
radiation, humidity and advection are absent. In the confounding structure of
Sec.~\ref{sec:numerics}, an omitted driver correlated with the true one transfers
attribution to whatever remains, which would bias these ratios upward. Working
against this, $H$ is during a heatwave partly a \emph{consequence} of soil
desiccation rather than an independent driver, and conditioning on a mediator
removes part of the effect being attributed. We do not claim a net direction for
either site, and the provenance of the two products adds a further caveat to the
physical reading, as noted above.

Third, the uncertainty of the fit is substantial. The bootstrap ranges quoted
above, which hold the knots, the residual autocorrelation and the climatological
reference at their full-sample values, span a factor of two at the 2003 site's
observed peak and of six at $38\,^{\circ}$C at the 2010 site, where the
extrapolation adds a structural
uncertainty that the bootstrap does not capture.

Fourth, each event is represented by one grid point rather than by the spatial
average over an exceedance region, so the numbers carry a sampling variability we
have not quantified. Together with the conditioning of
Eq.~\eqref{eq:appestimand}, this makes the ratios reported here not comparable term
by term with published regional estimates, including those of
Ref.~\cite{Miralles2014} for the same two events. The decisive test is the one
described in Sec.~\ref{sec:outlook}.

\subsection{Relation to information-flow theory}

Nothing here supersedes the entropy transfer, which remains the natural
quantity when the question concerns the uncertainty of the target as a whole.
What the bridge identity \eqref{eq:T3} establishes is that a small entropy
transfer is not evidence of a weak contribution to an extreme, and that the two
quantities can be computed from a single fitted model and reported together.
Section~\ref{sec:repairs} extends the statement from one measure to a family.
Three repairs of a vanishing flow, proposed for different reasons, all weigh or
expand the response to variations of the source, which vanishes with the
physical coupling at saturation, so the distinction drawn by Eq.~\eqref{eq:T3} is
a property of the class of measures and not of the Shannon functional in
particular. The saturated regime also shows that the distinction is not between
entropy and probability as such. A probability-based attribution measured about
the ensemble mean is as blind there as the entropy flow, and what restores the
contribution is a stated reference (Sec.~\ref{sec:saturation}).

The relation to the freezing operation of the entropy transfer is more specific
than it first appears. The flux decomposition avoids freezing, and the price
is that it does not attribute. The adjoint attribution reintroduces a freeze, applied to
the anomaly of the source rather than to its evolution
(Sec.~\ref{sec:estimation}), and inherits nil causality with it. The freezing
step is therefore not incidental to Liang's construction. It is what turns a
description of coupling into a statement about what the source causes, and
here it had to be reintroduced to pass from the decomposition to an
attribution.

\subsection{What the method requires}
\label{sec:requirements}

The practical content of Sec.~\ref{sec:numerics} is the condition of
Sec.~\ref{sec:requires}, namely that the fitted generator be Markov in an
observed state that contains the drivers of the target, together with the
requirement that the reference of the intervention be stated. This places the
burden where it belongs. The attribution step performs no variable selection
and faithfully reports whatever perturbation direction it is given, so the
scientific work lies in the conditioning set, the state representation and the
choice of reference rather than in the estimator. In the confounding structure
tested in Sec.~\ref{sec:numerics}, spurious attribution to an inert but strongly
correlated candidate falls to about two per cent once the true driver is in the
drift model alongside it, so an irrelevant candidate that is merely correlated
with a true driver costs little. That is not a licence to add every measurable variable. A variable
that is itself a consequence of the source, such as the $850$-hPa temperature
during a heatwave, is a mediator and removes part of the effect being attributed,
one that is a consequence of the target can create a spurious dependence, and
nearly collinear drivers make the individual components unstable. The
conditioning set should follow a temporal causal ordering of the variables, with
drivers of the target entering its drift and variables it drives kept out. Even
then, changing one fitted component represents a physical intervention only to
the extent that the fitted equations are structural, that is, that the other
components would remain as fitted if it were changed, and replacing the soil
forcing, prescribing a soil state and adding irrigation water are different
interventions.

The error budget has components of different sign, which should be reported as
such. First-order truncation understates the effect of removing a source that
shifts a log-concave terminal law toward the threshold, and for other
perturbations its sign is not fixed (Table~\ref{tab:saturation}). Its size depends
on the geometry of the perturbation. For the perturbation of the testbed, which narrows the terminal
law, it costs twenty-two to twenty-eight per cent at $\varepsilon=0.25$, nearly
independently of base rate. For a shift, as in the application, it follows
Eq.~\eqref{eq:meanshift} and decreases toward the tail. In the testbed
constructions, hidden slow memory and multiplicative noise act in the other
direction, by up to a factor of two, and state-dependent coupling in the same
direction, but these directions depend on the signs of the couplings and are not
general. An attribution computed under a model whose relaxation timescale is
physically implausible, whose residuals are autocorrelated or visibly
heteroscedastic, or whose hindcasts are miscalibrated should be treated as
unreliable in an unknown direction.

\subsection{Scope}
\label{sec:scope}

Three restrictions define what has been established. First, the drift is
source-additive, Eq.~\eqref{eq:additive}. A source entering multiplicatively
requires the interaction to be carried in the fitted model,
Sec.~\ref{sec:numerics} quantifies the cost of neglecting it, and
Sec.~\ref{sec:outlook} describes how the freezing construction of
Sec.~\ref{sec:estimation} could remove the restriction. Second, the response is
first order in the perturbation, so the attribution of a large intervention is
an extrapolation whose accuracy we have characterised (Secs.~\ref{sec:functional}
and~\ref{sec:saturation}) but not extended. Third,
the estimand is the probability of instantaneous exceedance at a fixed
threshold rather than of a persistent episode.

A fourth restriction deserves emphasis because it is easily mistaken for a
measurement of zero. The perturbation field \eqref{eq:perturbfield} acts on the
drift and holds the diffusion fixed. A drift perturbation can still change the
spread of the terminal law, as the ensemble-mean perturbation of the testbed
does, but a source that acts only through the target's noise amplitude receives
no attribution here, while remaining visible to a conditional mutual information,
which is sensitive to the full conditional law. In the intended application this
is not a remote possibility, because a dry soil lacks the moisture buffer that
damps the surface energy balance and so plausibly increases the variability of
temperature as well as its mean. The direction of the omitted effect is not fixed,
since a wider terminal law raises the probability of exceeding a threshold above
its mean and lowers it below, and several of the 2010 thresholds lie below the
mean. The ratios of Sec.~\ref{sec:application} therefore attribute one pathway of
a mechanism that has two, and they are neither an upper nor a lower bound on the
total. Extending the perturbation to the diffusion adds a term
$\tfrac12\,\delta\bm D:\nabla\nabla\psi$ to the integrand of
Eq.~\eqref{eq:response}, which could be evaluated and reported separately.

A fifth restriction concerns what the source variable must be. The
perturbation field acts on the drift of a state, so the framework attributes to
a source whose \emph{state} mediates the effect on the target. Where the effect
is mediated instead by a flux, a supply consumed as fast as it arrives, the
state carries no signature of the intervention, and the attribution returns a
null that reflects a failure of the observable rather than an absence of
effect. Irrigation is the obvious instance, since an irrigated crop can
transpire the applied water about as fast as it arrives, but the restriction applies to any
source that is supplied and used without accumulating, and it argues for
expressing such interventions as an input rate in the drift of the source
rather than as a displacement of its state.

\subsection{Outlook}
\label{sec:outlook}

The source-additive restriction can be lifted by defining the perturbation as a
difference of drifts under freezing, $\delta F = F|_{x_j\,\mathrm{frozen}} - F$,
which requires no additive structure, leaves the response identity
\eqref{eq:response} unchanged and retains nil causality by the same argument.
What it costs is the closed form. The perturbation is no longer read off a
per-driver fit, the freeze reference has to be specified explicitly rather than
read off the fitted component, and the operating envelope of
Sec.~\ref{sec:numerics} would have to be re-established for the wider model
class. That is the natural route to attribution under multiplicative or
interacting coupling, and it borrows the operation from Liang \cite{Liang2016}
rather than inventing one.

The persistence restriction is the least binding, although lifting it needs
more than a change of terminal condition. With absorbing conditions on a set,
$\psi$ becomes a first-passage probability or, on an infinite horizon, the
committor of transition-path theory \cite{EVandenEijnden2006}, and a mean
first-passage time solves a Poisson problem instead. The response identity
\eqref{eq:response} then keeps its form, with the time integral restricted to
paths not yet absorbed. Events defined by a duration above a threshold, which is
how heatwaves are defined in practice \cite{Seneviratne2010,Miralles2019},
additionally require the time already spent above the threshold as a state
variable, after which the same construction applies to the augmented Markov
process. A multivariate exceedance set
generalises Eq.~\eqref{eq:T1} to a surface integral of the current over the set
boundary, which is the natural route to compound events.

Every counterfactual against which the attribution has been checked was
computed inside a system we specified, so what has been shown is how accurate the
first-order truncation is, not that a fitted generator represents the system it
was fitted to. Whether an attribution computed from a fitted generator predicts
the effect of a real intervention is the claim that would matter most, and this
paper does not make it. The claim is, however, directly testable. Prescribed
and interactive land-surface experiments in climate models are executed
interventions in systems that contain the confounding, memory and forcing the
testbed lacks. Computing the adjoint attribution in the interactive
configuration alone, with an intervention and a conditioning that match those of
the prescribed experiment, and comparing it against the difference in event
frequency between configurations would establish or refute the proposition, and we regard
that as the decisive test of the approach. Paired irrigated and rainfed fields
would offer an observational counterpart, provided that the
intervention is expressed through the applied water rather than the soil-water
state (Sec.~\ref{sec:scope}).

\section*{Data availability}
ERA5 \cite{Hersbach2020} and ERA5-Land \cite{MunozSabater2021} are freely
available from the Copernicus Climate Change Service. The scripts that retrieve
the site series, run the testbed experiments and the attribution, and draw the
figures are openly available at \url{https://github.com/Dannyhay/pre-extremes-attribution}, where the version used for this paper, with the numerical output behind every figure and table, is tagged v1.0.

\begin{acknowledgments}
This work was supported by the European Research Council (ERC) under grant agreement 101088405 (HEAT) and the Research Foundation Flanders (FWO) under the FWO-NSFC bilateral research project funding (CausalHeat, G0AOO25N). Claude (Opus 5.5; Anthropic) was used in the course of this work to assist with analytic derivations and code writing for the experiments and figures. Every analytic result reported here was checked independently by the author, and every numerical result was reproduced from the scripts archived with the paper (see Data availability). The author takes full responsibility for the content.
\end{acknowledgments}

\appendix

\section{The information flow under source-additive drift}
\label{app:bridge}

Consider the system \eqref{eq:sde} with target drift $F_Y=f(Y,\bm Z,t)+b_tX$ and a
target diffusion $D_{YY}$ that does not depend on $X$. Let $\rho_t(y)$ be the
marginal density of $Y$ and $\rho_{\setminus X}(y,\bm z,t)=\int\rho\,dx$ the joint
density of all coordinates except the source. The Liang information flow from $X$
to $Y$ is \cite{Liang2008,Liang2016}
\begin{align}
 T_{X\to Y}=&-\E\!\left[\frac{1}{\rho_t}\int\partial_y\!\left(F_Y\rho_{\setminus X}\right)d\bm z\right]
 \nonumber\\
 &+\frac12\,\E\!\left[\frac{1}{\rho_t}\int\partial_y^2\!\left(D_{YY}\rho_{\setminus X}\right)d\bm z\right],
 \label{eq:liangflow}
\end{align}
where the expectations are over the joint law of $(X,Y)$. The integral over
$\bm z$ of the first bracket is $\partial_y\!\int f\rho_{\setminus X}\,d\bm z
+b_t\,x\,\partial_y\rho_t$, because $\partial_y$ acts at fixed $x$ and
$\int\rho_{\setminus X}\,d\bm z=\rho_t$. Its first part depends on $y$ only, and
its contribution to the expectation is $-\int\partial_y(\int f\rho_{\setminus
X}d\bm z)\,dy=0$ under the decay condition of Sec.~\ref{sec:flux}. Its second part
contributes $-b_t\int m_t\,\partial_y\rho_t\,dy$ with $m_t(y)=\E[X\mid Y=y]$, and
integrating by parts gives $b_t\int\rho_t\,\partial_y m_t\,dy$. The second term of
Eq.~\eqref{eq:liangflow} depends on $y$ only when $D_{YY}$ does not depend on $X$,
and its expectation is then the integral over $y$ of a second derivative, which
vanishes. Hence
\begin{equation}
 T_{X\to Y}=\int b_t\,m'_t\,\rho_t\,dy=\E\!\left[\Dresp(Y,t)\right],
 \quad \Dresp=b_t\,m'_t ,
\end{equation}
which is the local response of Sec.~\ref{sec:bridge} and, with
$\Dresp=\partial_y j$, Eq.~\eqref{eq:T3}. The further coordinates $\bm Z$ enter
only through $f$ and drop out, so the conditional mean is taken given $Y$ alone,
as in the specific source current. For a linear Gaussian system $m_t$ is linear
and $T_{X\to Y}=b_t\operatorname{Cov}(X,Y)/\Var(Y)$, Liang's multivariate linear
result \cite{Liang2021}.

\section{Derivation of the response identity}
\label{app:girsanov}

We derive Eq.~\eqref{eq:response} by a change of measure, following the martingale route of Ref.~\cite{GobetMunos2005}, for a fixed initial law at $t_0$, a fixed terminal event and coefficients regular enough for the backward equation \eqref{eq:backward} to have a solution with integrable gradient. Assume the
perturbation lies in the range of the diffusion, that is
\begin{equation}
 \delta\bm F(\bm x,t)=\bm B(\bm x,t)\,\bm h(\bm x,t)
 \label{eq:rangecond}
\end{equation}
for some bounded adapted $\bm h$, which guarantees the Novikov condition
$\E\exp\!\big(\tfrac12\int_{t_0}^{t_1}\!|\bm h|^2dt\big)<\infty$. This is the
condition under which the perturbed drift can be absorbed into a shift of the
driving Brownian motion. It holds when $\delta\bm F$ is bounded and the target
has noise of its own, independent of the other components and with an amplitude
bounded away from zero, as in the testbed. That the target is noisy is not by
itself sufficient, since $\bm B=(1,1)^{\top}$ forces both
coordinates of a two-dimensional system but cannot produce a perturbation of the
target's drift only. A perturbation outside the range of $\bm B$ requires the
alternative argument noted at the end of this appendix. The coloured temperature
noise of the application falls outside this argument, and there the response is
obtained directly as the derivative of the conditionally Gaussian terminal law of
Sec.~\ref{sec:estimation}, which is exact for the discrete model simulated.

Let $\mathbb P$ be the law of Eq.~\eqref{eq:sde} on $[t_0,t_1]$ and
$\mathbb P^{\varepsilon}$ that of the process with drift
$\bm F+\varepsilon\,\delta\bm F$ and the same diffusion. By Girsanov's theorem
the two are equivalent with
\begin{equation}
 \frac{d\mathbb P^{\varepsilon}}{d\mathbb P}
 = \exp\!\left(\varepsilon\!\int_{t_0}^{t_1}\!\bm h_t\cdot d\bm W_t
 - \frac{\varepsilon^{2}}{2}\!\int_{t_0}^{t_1}\!|\bm h_t|^{2}dt\right)
 \equiv Z_{\varepsilon}.
 \label{eq:rn}
\end{equation}
Since the observable is unchanged,
$P_u^{\varepsilon}(t_1)=\E_{\mathbb P}\!\left[\chi_u(Y_{t_1})\,Z_{\varepsilon}\right]$,
and differentiating at $\varepsilon=0$ with $Z_0=1$ gives
\begin{equation}
 \left.\frac{dP_u(t_1)}{d\varepsilon}\right|_{0}
 = \E\!\left[\chi_u(Y_{t_1})\!\int_{t_0}^{t_1}\!\bm h_t\cdot d\bm W_t\right].
 \label{eq:diffrn}
\end{equation}

The stochastic integral is removed by martingale representation. Because
$\psi(\bm x,t)=\E[\chi_u(Y_{t_1})\mid\bm U_t=\bm x]$ solves
Eq.~\eqref{eq:backward}, the process $\psi(\bm U_t,t)$ is a $\mathbb P$-martingale
and It\^o's formula leaves only its martingale part,
\begin{equation}
 d\psi(\bm U_t,t)=\left[\bm B^{\top}\nabla\psi\right](\bm U_t,t)\cdot d\bm W_t .
\end{equation}
Integrating from $t_0$ to $t_1$ and using $\psi(\cdot,t_1)=\chi_u$,
\begin{equation}
 \chi_u(Y_{t_1})=\psi(\bm U_{t_0},t_0)
 +\int_{t_0}^{t_1}\!\left[\bm B^{\top}\nabla\psi\right]\cdot d\bm W_t .
 \label{eq:mrt}
\end{equation}
Substituting Eq.~\eqref{eq:mrt} into Eq.~\eqref{eq:diffrn}, the term involving
$\psi(\bm U_{t_0},t_0)$ is deterministic given $\bm U_{t_0}$ and multiplies a
mean-zero It\^o integral, so it drops. The It\^o isometry then gives
\begin{align}
 \left.\frac{dP_u(t_1)}{d\varepsilon}\right|_{0}
 &= \int_{t_0}^{t_1}\!\E\!\left[
 \left(\bm B^{\top}\nabla\psi\right)\cdot\bm h\right]dt \nonumber\\
 &= \int_{t_0}^{t_1}\!\E\!\left[
 \left(\bm B\bm h\right)\cdot\nabla\psi\right]dt
 = \int_{t_0}^{t_1}\!\E\!\left[\delta\bm F\cdot\nabla\psi\right]dt ,
\end{align}
using Eq.~\eqref{eq:rangecond} in the last step. This is
Eq.~\eqref{eq:response}. The expectation is taken under the unperturbed law, so
the occupied distribution entering the estimator is the observed one.

When Eq.~\eqref{eq:rangecond} fails, the same identity follows by differentiating
the Kolmogorov equations directly, provided the forward density and $\psi$ are
regular enough for the duality between them to hold. Perturbing the generator by
$\delta\Lgen=\delta\bm F\cdot\nabla$ and applying that duality gives
$\delta P_u(t_1)=\int\langle\delta\Lgen^{\dagger}\rho_t,\psi\rangle\,dt
=\int\langle\rho_t,\delta\Lgen\psi\rangle\,dt$, which is the same expression. A
perturbation of the diffusion adds $\tfrac12\,\delta\bm D:\nabla\nabla\psi$ to
$\delta\Lgen\psi$, and an intervention that also changes the initial law adds
$\int\delta\rho_{t_0}\,\psi(\cdot,t_0)$. The measure-theoretic route is given here
because it makes the required regularity explicit.

\section{Small-$\alpha$ estimation of the R\'enyi transfer entropy}
\label{app:renyi}

A practical caution accompanies Eq.~\eqref{eq:T4}. As $\alpha\to0$, $Z_\alpha$
approaches the number of cells on which an estimated distribution is supported,
so a histogram estimate of $H_\alpha$ approaches the logarithm of the
occupied-cell count and depends on the support of the sample rather than on its
probabilities. We apply the binned R\'enyi transfer entropy in the sum form
given below, with the source forcing $g(S_t)$, the target $T_{t+\tau}$ and the
conditioning target $T_t$ at a lag $\tau$ of one sampling interval ($0.25$~d),
equidistant bins over the sample range (eight per variable unless stated), sixty
circular-shift surrogates of the source with shifts drawn uniformly from $[10,N-10]$
samples, and fifteen indices $\alpha$ from $0.1$ to $3$. Applied to the testbed of
Sec.~\ref{sec:numerics} with $N=2\times10^{5}$ samples, the raw statistic at
$\alpha=0.1$ orders the three regimes as wet $>$ very dry $>$ transitional,
inverting the ordering given by every other diagnostic and placing first the
regime in which the coupling is weakest by both level and slope. The values
also move with the discretisation, the very dry regime returning $0.082$,
$0.223$ and $0.195$ at five, eight and twelve bins per variable.

Against circular-shift surrogates, however, the regimes are ordered
transitional $>$ wet $>$ very dry at every $\alpha$ tested, which is the
ordering given by $T^{(\alpha)}$ and not the one given by the raw statistic.
The very dry regime reaches $|z|=0.67$ at its most favourable $\alpha$ and is
never significant. The apparent small-$\alpha$ signal is reproduced by the null
and is an artefact of occupancy. Surrogate testing is therefore essential at
small $\alpha$, as Ref.~\cite{Tabachova2026} also finds, where the raw statistic
is least interpretable, and once it is applied the estimator orders the regimes as
$T^{(\alpha)}$ does. The lagged statistic and $T^{(\alpha)}$ are different
functionals, so the agreement is one of ordering and not of value.

We record one further caution. The sum form
$H_\alpha(X,Z)+H_\alpha(Y,Z)-H_\alpha(X,Y,Z)-H_\alpha(Z)$ is not a proper
conditional mutual information for $\alpha\neq1$ and is not sign-definite. In
our runs the statistic and a few per cent of its surrogate draws take negative
values at $\alpha\gtrsim1.8$, so a one-sided significance test built on an
assumption of non-negativity is unsafe.

\bibliography{refs}
\end{document}